\documentclass[pre,twocolumn,10pt,aps,longbibliography,nofootinbib]{revtex4-1}

 \usepackage[utf8]{inputenc}
 \usepackage[greek,english]{babel}
 \usepackage{bm}
 \usepackage{verbatim}
 \usepackage{amsmath}
 \usepackage{amssymb}
 \usepackage{amsthm}
 \usepackage{latexsym}
 \usepackage{amsfonts}
 \usepackage{epsfig}
 \usepackage{epstopdf}
 \usepackage{wasysym} 
 \usepackage{subfigure}
 \usepackage{color}
 \definecolor{darkblue}{rgb}{0,0,.5}
 \definecolor{BLUE}{rgb}{0,0,1}
 \definecolor{BLACK}{rgb}{0,0,0}
 \usepackage[linktocpage, colorlinks=true, linkcolor=darkblue, citecolor=darkblue]{hyperref}
 \usepackage[all]{hypcap}
 \usepackage[makeroom]{cancel}

\newcommand{\C}[1]{{\cal{#1}}}
\newcommand{\bb}[1]{\textbf{#1}}
\newcommand{\mf}[1]{{\mathfrak{#1}}}
\newcommand{\lr}[1]{{\langle {#1}\rangle}}

\newcommand{\rl}[0]{{\rangle\langle}}
\newcommand{\blue}[1]{#1}
\newcommand{\new}[1]{#1}

\begin{document}

\title{Everything Everywhere All At Once: \\\blue{First Principles Numerical Demonstration of Emergent Decoherent Histories}}

\author{Philipp Strasberg}
\author{Teresa E. Reinhard}
\author{Joseph Schindler}
\affiliation{F\'isica Te\`orica: Informaci\'o i Fen\`omens Qu\`antics, Departament de F\'isica, Universitat Aut\`onoma de Barcelona, 08193 Bellaterra (Barcelona), Spain}

\date{\today}

\begin{abstract}
 Within the histories formalism the decoherence functional is a formal tool to investigate the emergence of classicality in isolated quantum systems, yet an explicit evaluation of it from first principles has not been reported. We provide such an evaluation for up to five-time histories based on exact numerical diagonalization of the Schr\"odinger equation. We find a robust emergence of decoherence for slow and coarse observables of a generic random matrix model and extract a finite size scaling law by varying the Hilbert space dimension over four orders of magnitude. Specifically, we conjecture and observe an exponential suppression of coherent effects as a function of the particle number of the system. This suggests a solution to the preferred basis problem of the many worlds interpretation (or the set selection problem of the histories formalism) within a minimal theoretical framework---without relying on environmentally induced decoherence, quantum Darwinism, Markov approximations, low-entropy initial states or ensemble averages.
\end{abstract}

\maketitle

\newtheorem{lemma}{Lemma}[section]

\section{Introduction}
\label{sec intro}

Do we live in a quantum Multiverse? Since the formulation of the many worlds interpretation (MWI)~\cite{EverettRMP1957, DeWittPT1970, Vaidman2021}, this idea enjoys increasing popularity among researchers~\cite{Carr2007, SaundersEtAlBook2010, WallaceBook2012}, popular science~\cite{CarrollPodcast2022} and pop culture~\cite{EEAAO2022}. Although the basic premise of the MWI is ``simply'' to assume \new{unitary evolution} for the entire Universe (including its observers), its seemingly absurd consequence that the Universe consists of many universes existing in parallel (the Multiverse) is a source of strong controversies~\cite{GisinArXiv2022}. Using non-relativistic quantum mechanics, our contribution is to provide direct evidence---within a minimal first principles setup---that the MWI is compatible with our experienced ``classical reality''.

It is useful to clarify from the start that the idea of the quantum Multiverse and the MWI should be distinguished from other potential Multiverses based on, e.g., \new{an infinitely large Universe}, many inflationary bubbles, different fundamental constants or various solutions in the string theory landscape~\cite{TegmarkSA2003, Carr2007}. Those universes are thought of being \emph{separated} in spacetime, and thus admit in principle a classical understanding. In contrast, the Multiverse of the MWI consists of different universes at the \emph{same} spacetime location.

\new{One} essential technical problem associated to the MWI is called the preferred basis problem: how to reconcile the Multiverse with our perceived classical experience within one universe? If \new{Schr\"odinger's equation} is applied to the entire Universe, then by linearity superpositions proliferate and spread, splitting the wave function into many universes, or histories, evolving in parallel (also called branches, worlds, realities, narratives, etc.). However, the wave function can---without approximation---be split with respect to many different bases (indeed, a continuum of bases) and each basis provides \emph{a priori} an equally justified starting point (and as we will discuss the wave function can be also split backwards in time). But as Bohr, in endless discussions with Einstein about the double slit experiment (and others), has made clear, a state describing a superposition of different properties makes said property \emph{ontologically} indeterminate: in the double slit experiment the particle has no meaningful location without measurement~\cite{Barad2007}. Thus, without identifying an \emph{additional structure} justifying a classical description, as we experience it, the MWI describes infinitely many ontologically indeterminate splittings: \emph{a priori} none of them allows to speak about ``histories'', ``worlds'' or ``realities'' in any conventional, meaningful sense.

This additional structure must be derived \new{within the MWI}, and within non-relativistic quantum mechanics a satisfactory derivation must comply at least with two minimal desiderata. (A) The system is isolated, evolves unitarily and is prepared in a pure state. This avoids the introduction of any form of classical noise from the outside (e.g., in form of ensemble averages), which potentially implies the answer to the question already from the start. We remark that the system might be (but does not need be) split into subsystems. (B) The condition of classicality must be a meaningful, rigorous definition that is suitable to account for \emph{multi-time} properties or \emph{temporal} correlation functions because the perception of classicality is a \emph{repeated experience} (adapting Einstein's quote, the moon was there yesterday, is there today and will be there tomorrow). This is important: speaking of different worlds or histories becomes meaningful if we can reason about their past, present and future in classical terms.

Here, we use the decoherence functional (DF) introduced within the consistent or decoherent histories framework (or simply ``histories framework'' for short)~\cite{GriffithsJSP1984, GellMannHartleInBook1990, OmnesRMP1992, DowkerHalliwellPRD1992, GellMannHartlePRD1993, HalliwellANY1995, DowkerKentJSP1996, GriffithsBook2002, GellMannHartlePRA2007, Griffiths2019} \new{to} rigorous\new{ly investigate the emergence} of multi-time classicality (see Sec.~\ref{sec frame} for the technical details). We numerically show that bases defined by a slow and coarse observable of a non-integrable system \new{are robustly decoherent according to this criterion}. More specifically, we show for up to five-time histories and a Hilbert space dimension $D$ varying over four orders of magnitude that quantum effects (suitably quantified below) are \emph{exponentially suppressed} as a function of the particle number $N$ of the system (with $N\sim\log D$). This provides a firm starting point to discuss the MWI \new{and to address the set selection problem of the histories framework~\cite{PazZurekPRD1993, DowkerKentPRL1995, DowkerKentJSP1996, RiedelZurekZwolakPRA2016}.} \blue{Our results also explicitly show} that the emergence of classicality is ubiquituous, that almost all initial wave functions \new{\emph{can}} give rise to interesting universes, and that the branching of the wave function is \emph{a priori} not related to any arrow of time.

Perhaps surprisingly, our results do \emph{not} rely on environmentally induced decoherence (EID)~\cite{ZurekRMP2003, JoosEtAlBook2003, SchlosshauerPR2019}\footnote{Note the unfortunate double meaning of decoherence in the literature:\emph{a priori} there is no connection between the DF in the histories framework and the concept of EID.} or its refinement to quantum Darwinism~\cite{ZurekNP2009, KorbiczQuantum2021, ZurekEnt2022}. This is surprising because the widely proclaimed (sole) answer to the question why the MWI gives rise to classically looking universes is EID, and many researchers worked on connecting the MWI or the histories framework to EID and quantum Darwinism~\cite{FinkelsteinPRD1993, SaundersFP1993, PazZurekPRD1993, DiosiEtAlPRL1995, BrunPRL1997, YuPA1998, VaidmanISPS1998, BrunPRA2000, RiedelZurekZwolakPRA2016, RiedelPRL2017,  AlbrechtBaunachArrasmithPRD2022, TouilEtAlArXiv2022, StrasbergSP2023}, yet an explicit evaluation of the DF following the desiderata (A) and (B) is still missing. Only if one invokes a rigorous notion of multi-time quantum Markovianity~\cite{PollockEtAlPRL2018, LiHallWisemanPR2018, MilzModiPRXQ2021}, clear connections between the DF and EID have been established~\cite{PazZurekPRD1993, DiosiEtAlPRL1995, BrunPRL1997, YuPA1998, BrunPRA2000, StrasbergSP2023}, but this only shifts the problem of proving multi-time \new{decoherence} to proving multi-time Markovianity, which is a daunting task too~\cite{DuemckeJMP1983, FordOConnellPRL1996, FigueroaRomeroModiPollockQuantum2019, FigueroaRomeroPollockModiCP2021, StrasbergEtAlPRA2023}.

In contrast, the present approach does not rely on any system-environment tensor product splitting, although it is important to emphasize that it is \emph{not} in conflict with EID or quantum Darwinism when applied to it. Instead, we only consider slow and coarse observables of isolated, non-integrable quantum systems, similar to the approach taken by van Kampen in 1954~\cite{VanKampenPhys1954}. While the importance of slow and coarse (or quasi-conserved) observables \new{for the emergence of decoherence} has been \new{anticipated in the histories formalism~\cite{GellMannHartlePRD1993, BrunHalliwellPRD1996, HalliwellPRD1998, HalliwellPRL1999, CalzettaHuPRD1999, HalliwellPRD2003, GellMannHartlePRA2007, HalliwellInBook2010}, non-integrability has never been considered} a key factor by proponents of the histories, EID or quantum Darwinism framework: van Kampen's work \new{remained} ignored despite him emphasizing its importance for the quantum-to-classical transition~\cite{VanKampenEtAlPT2000, VanKampenAJP2008}. \blue{More precisely, non-integrability is here taken to mean that random matrix theory captures well the relevant dynamic behaviour. While this is known to be true in many situations~\cite{Wigner1967, BrodyEtAlRMP1981, BeenakkerRMP1997, GuhrMuellerGroelingWeidenmuellerPR1998, HaakeBook2010, DAlessioEtAlAP2016, DeutschRPP2018}, in the context of the quantum-to-classical transition rigorous evidence for van Kampen's idea only slowly accumulated recently~\cite{GemmerSteinigewegPRE2014, SchmidtkeGemmerPRE2016, NationPorrasPRE2020, StrasbergEtAlPRA2023, StrasbergSP2023}. Moreover}, it has not yet been considered in light of the MWI and \new{a first principles evaluation of the DF remains missing.}

To summarize, our objective is to subject the MWI and the question whether it is compatible with our perceived classical reality to a rigorous, quantitative test based on the minimal desiderata (A) and (B). To this end, we introduce the general theoretical framework to address this question in Sec.~\ref{sec frame}. Then, we present extensive numerical results for a heat exchange model (an archetypical example of a nonequilibrium process in thermodynamics) in Sec.~\ref{sec numerics}.  Section~\ref{sec consequences} discusses \blue{the set selection problem in view of our findings,} and we conclude and provide perpectives in Sec.~\ref{sec perspectives}.

\section{\new{General Framework}}
\label{sec frame}

\subsection{\new{Mathematical definitions and problem}}
\label{sec math def}

We consider an isolated quantum system with Hilbert space $\C H$ of dimension $D=\dim\C H$ and Hamiltonian $H$. The time evolution operator from $t_j$ to $t_k$ is denoted $U_{k,j} = e^{-iH(t_k-t_j)}$ ($\hbar\equiv1$) and the initial state is $|\psi(t_0)\rangle$. Furthermore, $\{\Pi_x\}_{x=1}^M$ denotes a complete set of $M$ orthogonal projectors satisfying $\sum_{x=1}^M \Pi_x = I$ (with $I$ the identity) and $\Pi_x\Pi_y = \delta_{x,y}\Pi_x$. They divide the Hilbert space into a direct sum of subspaces: $\C H = \bigoplus_{x=1}^M\C H_x$. The dimension of the subspace $\C H_x$ equals the rank of the projector $\Pi_x$ and is denoted $V_x = \mbox{tr}\{\Pi_x\} = \dim\C H_x$. Note that $D = \sum_{x=1}^M V_x$.

Next, we decompose the unitarily evolved state $|\psi(t_n)\rangle = U_{n,0}|\psi(t_0)\rangle$ by writing $U_{n,0} = U_{n,n-1}\cdots U_{1,0}$ and inserting identities at times $t_n > \dots > t_1 > t_0$:
\begin{align}
 |\psi(t_n)\rangle
 &= \sum_{x_n}\Pi_{x_n} U_{n,n-1}\cdots\sum_{x_1}\Pi_{x_1}U_{1,0}\sum_{x_0}\Pi_{x_0}|\psi(t_0)\rangle \nonumber \\
 &\equiv \sum_{\bb x}|\psi(\bb x)\rangle. \label{eq psi tot}
\end{align}
Here, we abbreviated $\bb x = (x_n,\dots,x_1,x_0)$, which we call a \emph{history} of \emph{length} $L=n+1$ in the following. Moreover, $|\psi(\bb x)\rangle = \Pi_{x_n} U_{n,n-1}\cdots\Pi_{x_1}U_{1,0}\Pi_{x_0}|\psi(t_0)\rangle$ is the (non-normalized) state conditional on ``passing through'' subspaces $\C H_{x_j}$ at times $t_j$. Note that this construction becomes identical to Feynman's path integral if we let $t_{j+1}-t_j\rightarrow0$ and consider projectors $\Pi_x$ in the position representation.

Finally, we introduce the \emph{decoherence functional}
\begin{equation}\label{eq decoherence functional}
 \mf{D}(\bb x;\bb y) \equiv \lr{\psi(\bb y)|\psi(\bb x)},
\end{equation}
which is a Hermitian $M^L\times M^L$ matrix. Moreover, the important \emph{decoherent histories condition} (DHC) is defined by the condition
\begin{equation}\label{eq decoherence condition}
 \mf{D}(\bb x;\bb y) = 0 \text{ for all } \bb x\neq\bb y.
\end{equation}

The central task of this work is to understand the conditions when the DHC is \emph{generically} satisfied. In particular, we will see that it is hard to \emph{strictly} satisfy Eq.~(\ref{eq decoherence condition}) for the situations we are interested in. Hence, we will \emph{quantitatively} study from first principles a suitable smallness condition $\mf{D}(\bb x;\bb y) \approx 0$ discussed in Sec.~\ref{sec approximate decoherence} below. We remark that this is a well-defined and non-trivial mathematical problem---independent of the physical meaning attached to the different objects to which we turn now.

\subsection{\new{Physical meaning and further terminology}}

We continue with physical clarifications, also related to the previous literature.

First, we call $\{\Pi_x\}_{x=1}^M$ a \emph{coarse-graining} and $x$ a \emph{macrostate}, which is conventional terminology in statistical mechanics. Indeed, we are interested in observables that humans can perceive, and those are necessarily coarse. Quantitatively, this means that the number of projectors is much smaller than the Hilbert space dimension: $M\ll D$. Note that this is also satisfied in any quantum experiment that relies on macroscopic detectors.

Next, we briefly clarify the meaning of the DHC, which has been already discussed in the literature~\cite{GriffithsJSP1984, GellMannHartleInBook1990, OmnesRMP1992, DowkerHalliwellPRD1992, GellMannHartlePRD1993, HalliwellANY1995, DowkerKentJSP1996, GriffithsBook2002, GellMannHartlePRA2007, Griffiths2019}.\footnote{There has also been some controversy about the precise mathematical formulation of the DHC (see, e.g., Ref.~\cite{DiosiPRL2004}), but it seems that Eq.~(\ref{eq decoherence condition}) is nowadays universally accepted~\cite{Griffiths2019}. } To this end, recall that according to our classical (Newtonian or pre-quantum) ontology the world out there is made up of entities with independent and well-defined properties that can be revealed in principle to arbitary precision, i.e., any uncertainty about the state of the world is entirely subjective or epistemic (for a broader discussion see, e.g., Ref.~\cite{Barad2007}). This worldview is challenged by quantum physics. However, the DHC guarantees that, \emph{for} the process describing the macrostates $x_j$ at times $t_j$, such a classical ontology becomes quantitatively accurate because Eq.~(\ref{eq decoherence condition}) describes the absence of detectable quantum interference effects. This makes the dynamics of the coarse properties $x_j$ isomorphic to a classical stochastic process (see also Refs.~\cite{SmirneEtAlQST2018, StrasbergDiazPRA2019, MilzEtAlQuantum2020, MilzEtAlPRX2020, MilzModiPRXQ2021, StrasbergBook2022, StrasbergEtAlPRA2023, StrasbergSP2023, SzankowskiCywinskiArXiv2023}) and implies the validity of Leggett-Garg inequalities~\cite{EmaryLambertNoriRPP2014}.

Before turning to the connection with the MWI, it is important to be clear about the fact that the DHC has differing ontological statuses even among practitioners of the histories formalism~\cite{OmnesRMP1992, DowkerKentJSP1996}. For instance, Griffiths regards the DHC as \emph{primary}, that is, quantum evolution is fundamentally stochastic and the deterministic Schr\"odinger equation can only be used to compute the DF \emph{if} the DHC is satisfied~\cite{GriffithsJSP1984, GriffithsBook2002, Griffiths2019}. In contrast, we here take unitary evolution as primary (as in the MWI) and view the DHC as an \emph{emergent} property. This means we use the histories formalism as a convenient mathematical tool to address a well-defined problem.

We deliberately point out that this attitude shall neither imply that the MWI is correct not that the consistent histories interpretation of Griffiths is incorrect. In contrast, the question we ask here sheds light on both interpretations. It is important to know for the MWI whether the quantum Multiverse supports decoherent histories, and likewise the question with which histories Schr\"odinger's equation is compatible, and whether this compatibility is exact or approximate, influences the scope of the consistent histories interpretation.

We continue by commenting on the connection between the DHC and the MWI as we view it in this work. As emphasized in the introduction, the definition of a ``world'' within the MWI is \emph{a priori} complicated by the existence of a continuum of mathematically conceivable different worlds (preferred basis problem). Moreover, even among the proponents of the MWI there is no universally agreed on quantitative definition of a world \blue{(see, e.g., Ref.~\cite{JessRiedelBlog2023} for a short overview)}. Here, we use this liberty and interpret the DHC as a minimal criterion to define a world within the MWI. This is motivated by what we said above: the classical world that we perceive is compatible with quantum mechanics precisely when the observables we look at satisfy the DHC. We believe this view is in unison with proponents of the MWI, for instance, Vaidman verbally defines a world as ``the totality of macroscopic objects [...] in a definite classically described state''~\cite{Vaidman2021}. Moreover, \blue{only past ``events'' giving rise to decoherent histories can leave records about having ``happened'' in the present state of the Universe, which highlights the importance of the DHC}~\cite{AlbrechtPRD1992, GellMannHartlePRD1993, FinkelsteinPRD1993, PazZurekPRD1993, HalliwellPRD1999, DoddHalliwellPRD2003, RiedelZurekZwolakPRA2016, HartleArXiv2016}.

In the following, it is also advisable to drop the word ``classical'' in favor of ``decoherent'', at least in a technical context. Indeed, depending on the context, classicality can have many different meanings since the boundary between quantum and classical physics can not be reduced to a single condition. Thus, in short, our terminology is the following: histories that satisfy exactly or approximately (see Sec.~\ref{sec approximate decoherence}) the DHC are called decoherent (and only sometimes classical) and in the context of the MWI we call those histories also branches or worlds. We also repeat once more that the DHC is not identical to but compatible with the concept of EID.

We conclude by specifying what we mean by a \emph{first principles} demonstration of the DHC. Essentially, we aim at a general understanding of the DHC by using physical assumptions that are not in conflict with the framework of Sec.~\ref{sec math def}. This has two major implications. First, we do not allow for any assumption that breaks unitarity and, instead, we solve the Schr\"odinger equation exactly. In particular, we do not use Markov approximations as done in Refs.~\cite{PazZurekPRD1993, DiosiEtAlPRL1995, BrunPRL1997, YuPA1998, BrunPRA2000, SmirneEtAlQST2018, StrasbergDiazPRA2019, MilzEtAlPRX2020, StrasbergSP2023}, which---even though insightful---shift the problem from justifying classicality to justifying Markovianity.

Second, we demand that the state of the isolated system is pure. By avoiding classical ensemble averages, any remaining uncertainty must stem from the quantum state or dynamics. Indeed, the only explicit evaluation of the DHC for a non-trivial many-particle system (and not relying on Markov approximations) has been done using the influence functional for a harmonic oscillator environment (Caldeira-Leggett model), \emph{assuming} that the environment is prepared in a canonical Gibbs ensemble~\cite{SchmidAP1987, DowkerHalliwellPRD1992, GellMannHartlePRD1993, HalliwellPRD1999, HalliwellPRD2001, SubasiHuPRE2012}. This is problematic as it remains unclear whether the observed decoherence is a consequence of the dynamics itself (as one would hope for) or a consequence of the classical ensemble average.

What remains beyond Markov approximations and ensemble averages, and beyond conserved quantities that trivially satisfy the DHC, are indirect arguments based, e.g., on the approximate behavior of projectors in the Heisenberg picture~\cite{OmnesJSP1989, OmnesRMP1992, GellMannHartlePRD1993, BrunHalliwellPRD1996, HalliwellPRD1998, HalliwellPRL1999, CalzettaHuPRD1999, HalliwellPRD2003, GellMannHartlePRA2007, HalliwellInBook2010}.\footnote{It is also worth to note that most of these indirect arguments aimed at establishing the DHC \emph{plus} the fact that there is only a single history happening (i.e., a deterministic history framework). This is more than what is necessary in our view (the existence of classical stochastic processes is an experimental fact). Moreover, somewhat confusingly this \emph{stronger} condition (DHC plus determinism) has been often called ``quasi-classicality'', which verbally suggests something \emph{weaker} than classicality as used here. This is another reason for us to prefer the word decoherence over the word classicality. } But an explicit evaluation of the DHC or even an estimate of it has never been attempted therein. It is this important gap that we fill with the present work, which is backed up by the general idea of van Kampen~\cite{VanKampenPhys1954} (see also Sec.~\ref{sec physical origin}) and supported only by a few preliminary results so far~\cite{GemmerSteinigewegPRE2014, SchmidtkeGemmerPRE2016, NationPorrasPRE2020, StrasbergEtAlPRA2023, StrasbergSP2023}.

\subsection{\new{Approximate decoherence}}
\label{sec approximate decoherence}

For reasons that will become clear below, exact decoherence, i.e., a strict satisfaction of Eq.~(\ref{eq decoherence condition}), is not the rule. All we can realistically hope for is to satisfy the DHC approximately, even though our results will also indicate that this approximation is typically so good that it becomes indistinguishable from exact decoherence for all practical purposes.

Approximate decoherence is typically quantified---after realizing $|\mf{D}(\bb x;\bb y)|\le\sqrt{\mf{D}(\bb x;\bb x)\mf{D}(\bb y;\bb y)}$ due to Cauchy-Schwarz---by introducing a ``normalized DF'' and demanding that~\cite{DowkerHalliwellPRD1992}
\begin{equation}\label{eq approximate decoherence}
 \epsilon(\bb x;\bb y) \equiv \frac{|\mf{D}(\bb x;\bb y)|}{\sqrt{\mf{D}(\bb x;\bb x)\mf{D}(\bb y;\bb y)}} \ll 1 \text{ for all } \bb x\neq\bb y.
\end{equation}
While this is a reasonable, properly normalized condition for smallness, in practice this condition can still be cumbersome. There are $M^{2L-1}-M^L$ many $\epsilon(\bb x;\bb y)$ for histories of length $L=n+1$ that are not trivially one because of $\bb x=\bb y$ or trivially zero because of $x_n\neq y_n$ (owing to the ortogonality of the projectors at the final time). Studying all of them becomes unfeasible for large $M$ or $L$. Moreover, the exact operational meaning of Eq.~(\ref{eq approximate decoherence}) is at least \emph{a priori} not apparent, i.e., what does it imply for an experimentalist who tries to decide whether a process is decoherent or not?

We therefore choose a dual strategy to quantify decoherence in this work. First, we consider the average of the non-trivial values of $\epsilon(\bb x;\bb y)$, defined as
\begin{equation}\label{eq epsilon avg}
 \epsilon \equiv \frac{1}{M^{2L-1}-M^L} \sum_{\bb x\neq\bb y} \epsilon(\bb x;\bb y).
\end{equation}
This is probably the simplest quantifier one can consider, but we believe its simplicity makes it appealing to get a first impression of what is going on.

\begin{figure}[t]
 \centering\includegraphics[width=0.32\textwidth,clip=true]{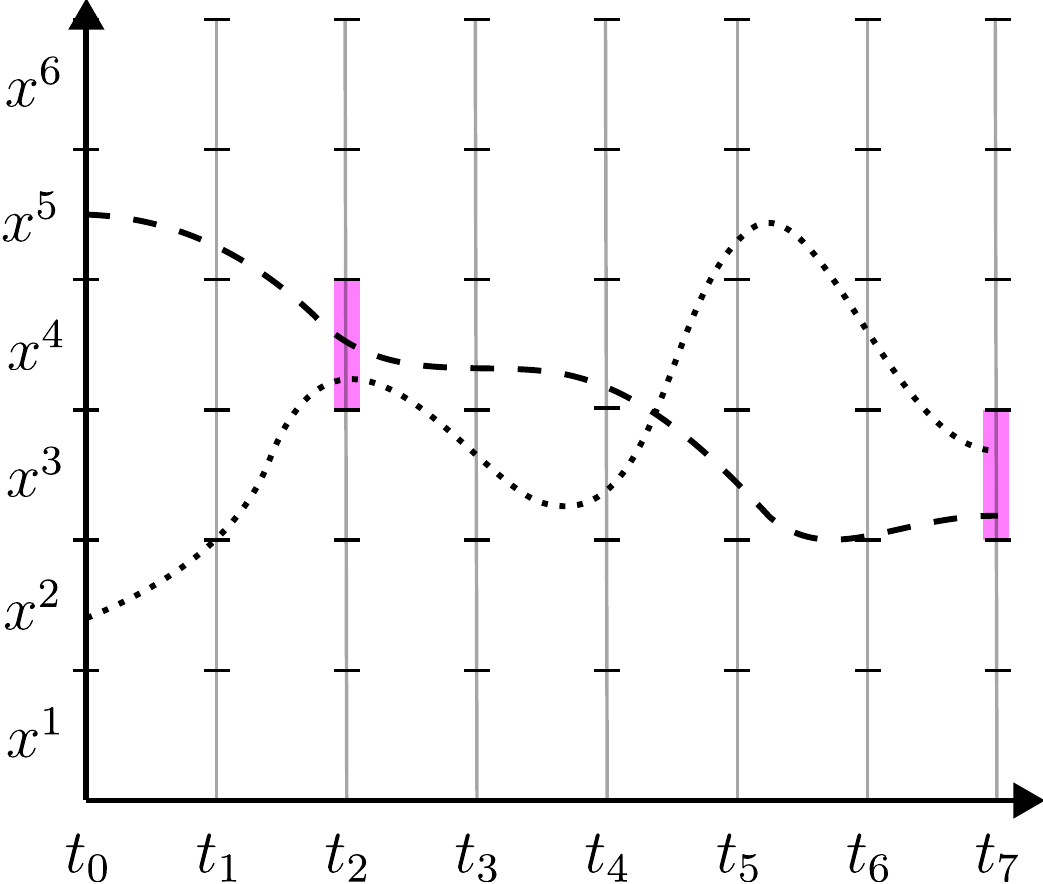}
 \label{fig paths}
 \caption{Example illustrating the DF for a coarse-graining with six macrostates and eight time steps. Two possibly interfering histories are indicated by dashed and dotted lines (the lines are used for visualization only, within this DF only their macrostates at times $t_i$ are defined). }
\end{figure}

However, to rigorously quantify the worst case scenario in an operationally meaningful way, we also consider a second strategy. To this end, we select an arbitrary subset $T\subset\{t_0,t_1,\dots,t_n\}$ of times on which the DF is defined, for instance, $T = \{t_2,t_7\}$ in the sketch of Fig.~\ref{fig paths}. Furthermore, we denote by $Z_T$ all possible histories $\bb z$ that we can associate to $T$ with respect to the given coarse-graining, e.g., with respect to Fig.~\ref{fig paths} these are all 2-time histories $Z_T = \{x^1_7,x^2_7,\dots,x^6_7\} \times \{x^1_2,x^2_2,\dots,x^6_2\}$ (with $\times$ the Cartesian product).

Next, we construct two types of probabilities on $Z_T$. The first is the actual Born rule probability to measure $z\in Z_T$, which we can get from the DF defined on $T$ via
\begin{equation}\label{eq prob quantum}
 p(\bb z) \equiv \sum_{\bb x(\bb z),\bb y(\bb z)} \mf{D}(\bb x;\bb y),
\end{equation}
where $\sum_{\bb x(\bb z)}$ indicates a sum running over all $\bb x$ holding $\bb z$ fixed. For instance, for the example in Fig.~\ref{fig paths} (the pink bars) we have $\sum_{\bb x(\bb z)} = \sum_{\bb x} \delta_{x_7,x^3_7} \delta_{x_2,x_2^4}$ and thus
\begin{equation}
 p(\bb z) = \lr{\psi(t_0)|U^\dagger_{2,0}\Pi_{x_4}U^\dagger_{7,2}\Pi_{x_3}U_{7,2}\Pi_{x_4}U_{2,0}|\psi(t_0)}.
\end{equation}
The second probability we can introduce is a decohered or classical version of $p(\bb z)$, obtained under the assumption that the DHC is obeyed. It is defined as a sum over the diagonal elements only:
\begin{equation}\label{eq prob classical}
 p_\text{cl}(\bb z) \equiv \sum_{\bb x(\bb z)} \mf{D}(\bb x;\bb x).
\end{equation}

Now, to evaluate the difference between the true quantum probability distribution $p(\bb z)$ and its decohered counterpart $p_\text{cl}(\bb z)$ we use the $L_1$-norm or trace distance
\begin{equation}\label{eq trace distance}
 \Delta_T(p|p_\text{cl}) \equiv \frac{1}{2}\sum_{\bb z\in Z_T} |p(\bb z)-p_\text{cl}(\bb z)| \in[0,1].
\end{equation}
Inserting the definitions, we see that the trace distance is determined by the off-diagonal elements of the DF:
\begin{equation}
 \Delta_T(p|p_\text{cl}) = \frac{1}{2}\sum_{\bb z\in Z_T} \left|\sum_{\bb x(\bb z)\neq\bb y(\bb z)} \mf{D}(\bb x;\bb y)\right|.
\end{equation}
In addition, the trace distance has a clear operational meaning in a hypothesis testing scenario~\cite{WildeBook2019}. Namely, it determines in an optimized single shot scenario the minimum probability $P_\text{min}(\text{fail}) = [1-\Delta_T(p|p_\text{cl})]/2$ for an experimenter to fail to distinguish between the coherent and decohered quantum process, described by $p(\bb z)$ and $p_\text{cl}(\bb z)$, respectively. For a given measurement resolution this gives the desired criterion to decide when a set of approximately decoherent histories is ``decohered enough''.

However, the trace distance~(\ref{eq trace distance}) still depends on $T$, but we are only interested in the worst case scenario. Moreover, we notice the following fact: if $t_n\notin T$, then the sum over $x_n$ and $y_n$ does not contribute to Eq.~(\ref{eq trace distance}). This follows from a \emph{containment} property of the DF: the DF for histories of length $L'<L$ can be obtained from the DF for histories of length $L$ by tracing out the final $L-L'$ time steps. Thus, we will plot in Sec.~\ref{sec numerics} the maximum
\begin{equation}\label{eq TD max}
 \Delta^\text{max}_L \equiv \max_{T\cap\{t_n\}\neq\emptyset} \Delta_T(p|p_\text{cl}).
\end{equation}
Note that this number still depends on $L$.

\subsection{\new{Physical origin of decoherence}}
\label{sec physical origin}

The physical origin of decoherence has been discussed at many places and shall not be repeated here in detail. We therefore limit ourselves to listing the four assumptions used by van Kampen~\cite{VanKampenPhys1954} followed by a perhaps oversimplified yet hopefully intuitive analogy.

The four assumptions are the following. First, one needs a large Hilbert space dimension $D$, which is certainly satisfied for macroscopic systems. Typically, $D=\C O(10^N)$ and $N=\C O(10^{23})$ to give some numbers. Second, one must consider a coarse coarse-graining satisfying $M\ll D$. As discussed above, this is satisfied for all that we humans can perceive. Third, the considered projectors must evolve slowly in a suitable sense such that the unobserved microscopic degrees of freedom have time to self-average or randomize. Common examples include hydrodynamic modes, collective degrees of freedom, weakly coupled subsystems, or more generally projectors that almost commute with the total Hamiltonian \blue{(for a more technical discussion about the notion of slowness see Ref.~\cite{StrasbergEtAlPRA2023})}. Fourth, the isolated system should be non-integrable.

This last point is certainly still subject to debate, not the least because a universally agreed on definition of quantum (non-)integrability is lacking. For instance, van Kampen assumed that the energy gap spectrum of the Hamiltonian $H$ is non-degenerate (apart from rare accidental degeneracies)~\cite{VanKampenPhys1954}, the results of Ref.~\cite{GemmerSteinigewegPRE2014, SchmidtkeGemmerPRE2016, NationPorrasPRE2020, StrasbergEtAlPRA2023} were based on the eigenstate thermalization hypothesis~\cite{DAlessioEtAlAP2016, DeutschRPP2018}, and here we use random matrix theory as also done in Ref.~\cite{NationPorrasPRE2020, StrasbergSP2023}. Since already classical systems tend to be chaotic (e.g., the Newtonian three-body problem), we believe that non-integrability can hardly count as an assumption in our Universe, even though it is an open question whether it is strictly necessary.

Intuitively, we like to motivate the need for non-integrability by the following analogy. Since the DHC describes the lack of interference between different macrostates $x$, we like to picture the coherences of all microscopic degrees of freedom as ripples caused by stones thrown into an initially still pond. If one zooms in very much (corresponding to very large $M$) or considers only a single stone thrown into the pond (corresponding to very small $D$), one can certainly see a clear wave pattern and interference effects. But if one zooms out a bit, or averages over a small area of the pond, and throws in a lot of stones (at different places), it becomes very hard to see any significant interference effects. One reason for it is simply the law of large numbers: there are many more combinations possible where ripples from different stones destructively interfere than possible combinations of constructive interference. This is identical to rolling $10^{23}$ dice: the vast majority of sequences has an average face value around 3.5 and the number of sequences deviating significanty from it is exponentially suppressed.

However, there is also another effect at work. In reality all stones have a slightly different size and weight, thus causing ripples with, e.g., different amplitudes or velocities. This implies that even if at a certain moment in time the state of the stones is highly synchronized (for instance, they could have been thrown symmetrically into the pond), this synchronized state will very quickly dephase and look generic. This effect is absent in integrable systems where the extensive amount of conserved quantities causes many regularities in the dynamics. It is thus reasonable to conjecture that non-integrabe systems, here corresponding to stones of different sizes and weights, show a more robust emergence of decoherence for a broader class of initial states.

\section{Numerical Demonstration}
\label{sec numerics}

\subsection{Setup}

We will now explicitly validate the idea that coarse and slow observables of a non-integrable many-body system satisfy 
the DHC with increasing accuracy for increasing Hilbert space dimension $D$. This is done by exact numerical 
diagonalization for a setup consisting of two identical, coupled systems exchanging energy. We choose this
archetypical setup of a nonequilibrium thermodynamic process because it is intuitive and allows us to clearly identify
arrow(s) of time associated with the flow of heat from hot to cold. In principle, this flow of heat could then be
harnessed to create work or free energy, an important prerequisite for the formation of life and intelligent observers
in the Universe, but (obviously) our code can not simulate any forms of life or observers.

\begin{figure}[t]
 \centering\includegraphics[width=0.42\textwidth,clip=true]{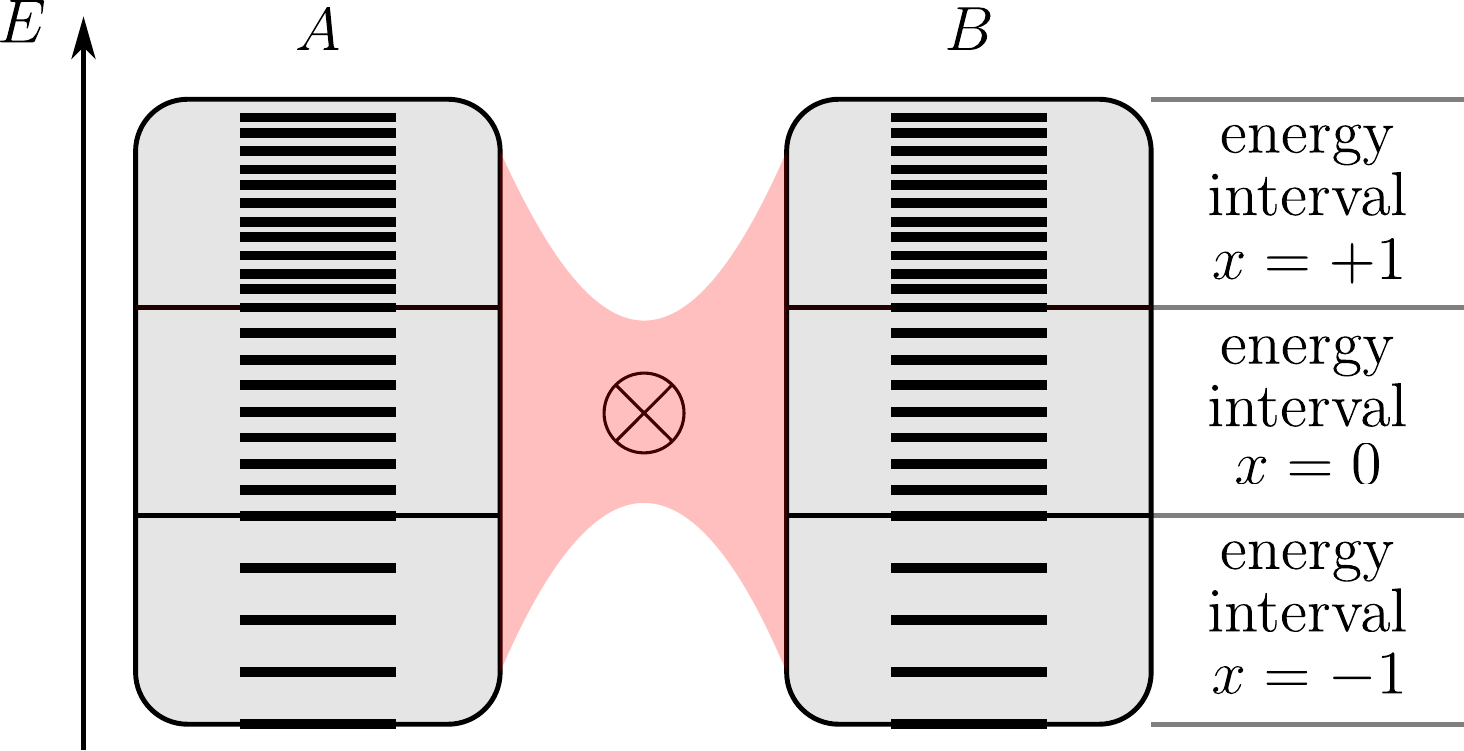}
 \label{fig sketch setup}
 \caption{Sketch of two identical, interacting systems $A$ and $B$ with discrete energy levels. The energies of both $A$ and $B$ are coarse-grained into windows $x$ with an increasing number of levels. Finally, the dynamics is restricted to a microcanonical subspace of the total energy corresponding to windows $(x_A,x_B) = (+1,-1) \cup (0,0) \cup (-1,+1)$.}
\end{figure}

We call our systems $A$ and $B$ (see Fig.~\ref{fig sketch setup} for a sketch) and the total Hamiltonian is
\begin{equation}\label{eq H mic}
 H = H_A + H_B + \lambda H_I,
\end{equation}
where $H_I$ is the interaction and $\lambda$ a small parameter to ensure weak coupling (see below). We coarse-grain the local energies of $A$ and $B$ by projectors $\Pi_{x,x'} \equiv \Pi_x^A\otimes\Pi_{x'}^B$ for a pair of integers $(x,x')$. In units of some energy precision or width $\Delta E$, $\Pi_{x,x'}$ projects on the energy interval $[x,x+1)\times[x',x'+1)$, i.e., $\Pi_x^A$ is spanned by all energy eigenstates $|k\rangle_A$ of $H_A$ whose energy eigenvalue $E_k$ satisfies $x\le E_k/\Delta E < x+1$, and similarly for $\Pi_{x'}^B$.

In the following, we consider one \emph{microcanonical subspace} with some fixed total energy 
$E_\text{tot} = E_A + E_B$, where $E_A$ ($E_B$) is the coarse-grained energy of system $A$ ($B$) determined with 
precision $\Delta E$. Thus, suppose we fix the total energy to be $E_\text{tot} = m\Delta E$ with $m\in\mathbb{Z}$, 
then the projector on this microcanonical subspace is given by
\begin{equation}
 P_{m} = \sum_{x,x'} \delta_{m,x+x'}\Pi_{x,x'}.
\end{equation}
The Hamiltonian restricted to this subspace is consequently given by
\begin{equation}\label{eq H rest}
 H_m \equiv P_{m} H P_{m} = \sum_{x,y} \Pi_{x,m-x} H \Pi_{y,m-y}.
\end{equation}

Specifically, we set (without loss of generality) in the following $E_\text{tot} = 0$. Moreover, as sketched in Fig.~\ref{fig sketch setup}, we consider for simplicity only three participating energy macrostates in $A$ and $B$, i.e., $P_0 = \Pi_{-1,+1} + \Pi_{0,0} + \Pi_{+1,-1}$. Of course, for a realistic macroscopic system more subspaces should be considered, but three subspaces are sufficient for a proof-of-principle demonstration of our main ideas.

Since the total energy is fixed, we write in the following $(\Pi_-,\Pi_0,\Pi_+) \equiv (\Pi_{-1,+1},\Pi_{0,0},\Pi_{+1,-1})$ with associated subspace dimensions $(V_-,V_0,V_+)$. Note that these projectors project on subspaces of systems $A$ \emph{and} $B$. The restricted Hamiltonian~(\ref{eq H rest}) has consequently nine blocks:
\begin{equation}\label{eq block H}
 H_0 = \begin{pmatrix}
        H_{--} & H_{-0} & H_{-+} \\
        H_{0-} & H_{00} & H_{0+} \\
        H_{+-} & H_{+0} & H_{++} \\
       \end{pmatrix}.
\end{equation}
Each of these blocks is in principle fully determined by the microscopic Hamiltonian~(\ref{eq H mic}), but we only want to capture four main features here. First, the systems $A$ and $B$ are assumed to be \emph{identical} from a thermodynamic point of view. In particular, we assume that the relation between energy and temperature is the same in $A$ and $B$ such that their energy difference is proportional to their temperature difference. Second, we consider \emph{normal} thermodynamic systems where the subspace corresponding to an equal energy distribution (or the same temperature according to the previous agreement) is the largest ``equilibrium'' subspace, i.e., $V_0 > V_-,V_+$. Third, we assume a \emph{weak interaction} between $A$ and $B$, i.e., the interaction energy $\lambda H_I$ is supposed to be negligible (otherwise it would not be justified to restrict the discussion to a microcanonical energy window defined by the sum $E_A+E_B$ of local energies only). Fourth, we want to mimic the interaction of two \emph{generic} complex (and thus non-integrable) many-body systems. According to common knowledge in statistical mechanics, this can be efficiently done by using random matrix theory~\cite{Wigner1967, BrodyEtAlRMP1981, BeenakkerRMP1997, GuhrMuellerGroelingWeidenmuellerPR1998, HaakeBook2010, DAlessioEtAlAP2016, DeutschRPP2018}.

In unison with these agreements we consider below the following specific Hamiltonian. The diagonal blocks
$(H_{--},H_{00},H_{++})$ are modeled by diagonal matrices with ($V_-,V_0,V_+$) many equally spaced energies in the
interval $[0,2\Delta E)$.\footnote{The factor ``2'' comes from the fact that the energy uncertainty of the projector
$\Pi_{x,x'} \equiv \Pi_x^A\otimes\Pi_{x'}^B$ is $2\Delta E$ because both $\Pi_x^A$ and $\Pi_{x'}^B$ have an energy
uncertainty $\Delta E$. Moreover, assuming $H_{--}$, $H_{00}$ and $H_{++}$ to be diagonal can always be achieved by a
block-unitary transformation. Finally, equal energy spacing is used for convenience only, a random spacing, for instance,
has been observed to give rise to the same behaviour.} Moreover, we set $H_{--} = H_{++}$, implying $V_-=V_+$ such that
the total Hilbert space dimension is $D=2V_- + V_0$. The coupling between different blocks is realized by random
matrices $H_{-0} = H_{0-}^\dagger$ and $H_{0+} = H_{+0}^\dagger$ with elements drawn from an othogonal
zero-mean-unit-variance Gaussian ensemble (the unitary Gaussian ensemble was not observed to give rise to different  behaviour) multiplied by the small coupling constant $\lambda$. Note that $H_{-0}$ and $H_{0+}$ are not square
matrices since $V_0 > V_-,V_+$. Moreover, in view of the weak interaction, we set $H_{-+} = H_{+-}^\dagger$ to be a
matrix of zeros, i.e., we forbid transitions between energy levels that are too far away from each other.

To remain in the regime of weak coupling, but to ensure that the different energy levels in $A$ and $B$ also sufficiently interact to make energy transport efficient, we have to choose $\lambda$, $\Delta E$ and the volumes appropriately. Their values can be estimated as follows. First, let us define weak coupling by demanding that the energy of the diagonal part $H_0 = H_{--} + H_{00} + H_{++}$ dominates the interaction energy $\lambda H_I = H_{-0} + H_{0-} + H_{+0} + H_{0+}$. To estimate their typical value we use an average over the microcanonical ensemble $\lr{\dots}_\text{mic}$. For the diagonal part we find the value $\lr{H_0}_\text{mic} \approx \Delta E$. Since $\lr{\lambda H_I}_\text{mic} = 0$, we look at the standard deviation $\lambda\lr{H_I^2}_\text{mic}^{1/2} \approx \lambda \sqrt{V_0V_\pm/D} \approx \lambda\sqrt{V_\pm}$. Thus, we obtain the condition $\lambda^2V_\pm/\Delta E^2 \ll 1$ (note that we take $\lambda$ to have the dimension of energy). To make $A$ and $B$ sufficiently interact we note that the interaction smears out the local energy eigenstates of $A$ and $B$ proportional to $\lambda$ (``level broadening''). At the same time, the density of states in the $x=\pm1$ subspace is roughly $\Delta E/V_\pm$. To guarantee that the smeared out levels in $A$ ($B$) overlap with sufficiently many levels in $B$ ($A$), we thus demand $(\lambda V_\pm/\Delta E)^2 \gg1$. Specifically, what we found to work numerically well are the conditions
\begin{equation}\label{eq conditions epsilon}
 \frac{1}{V_\pm}\left(\frac{\pi\lambda V_\pm}{2\Delta E}\right)^2 \ll 1, ~~~
 32\left(\frac{\pi\lambda V_\pm}{2\Delta E}\right)^2 \gg 1,
\end{equation}
compare also with Ref.~\cite{BartschSteinigewegGemmerPRE2008} for an analytical study of a similar model (two equal energy bands coupled via a random matrix). Unless otherwise mentioned, we choose $\lambda$ below such that the left equation reduces to $0.01$ $(\ll 1)$, which implies that the right equation is well satisfied for $V_\pm\gtrsim300$. Moreover, the characteristic evolution time scale has been found to be well approximated by
\begin{equation}\label{eq tau}
 \tau = \frac{\Delta E}{4\pi\lambda^2V_\pm}.
\end{equation}
In all that follows we set $\Delta E = 1$ and $V_0 = 3V_-$.

Finally, we write the initial state as
\begin{equation}\label{eq initial state}
 |\psi(t_0)\rangle = \sqrt{p_-(0)}|\psi_-\rangle + \sqrt{p_0(0)}|\psi_0\rangle + \sqrt{p_+(0)}|\psi_+\rangle.
\end{equation}
Here, $p_x(0)$ with $x\in\{-,0,+\}$ is the (\emph{a priori} arbitrary) probability to find the system in macrostate $x$. Unless otherwise mentioned, the states $|\psi_x\rangle$ are normalized Haar randomly chosen states in the subspace $\C H_x$ on which $\Pi_x$ projects. By choosing them Haar randomly we guarantee an unbiased choice, somewhat in spirit of a maximum entropy principle for pure states. In particular, given that $p_x(0)$ is the only available information, this allows us to ask about the typical behaviour of the system.

\subsection{Results}

\begin{figure}[tb]
 \centering\includegraphics[width=0.47\textwidth,clip=true]{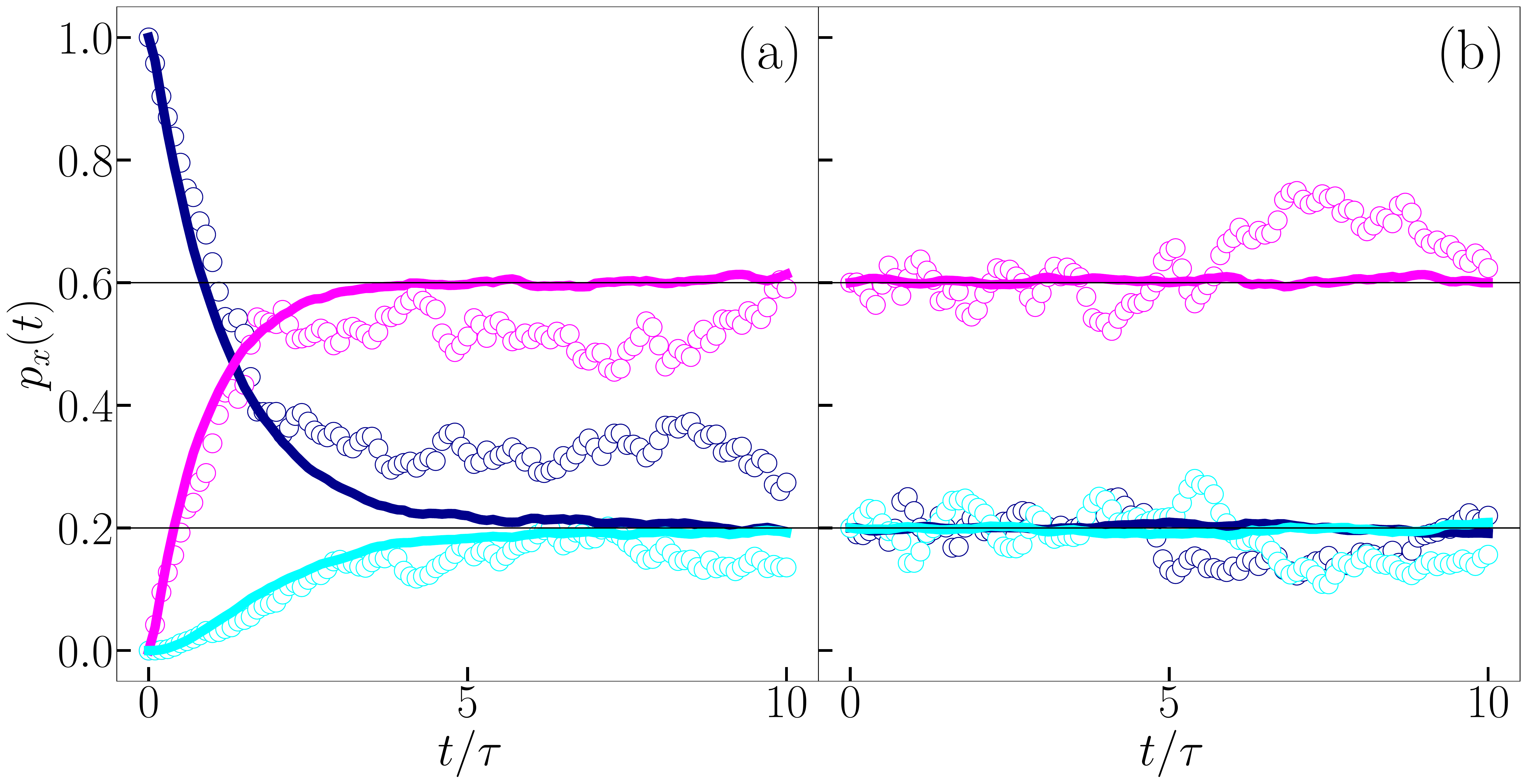}
 \label{fig dynamics}
 \caption{Time evolution of $p_-(t)$ (dark blue), $p_0(t)$ (magenta) and $p_+(t)$ (cyan) over dimensionless time
 $t/\tau$ for an initial nonequilibrium state (a) or equilibrium state (b) for $D=10^4$ (solid lines) and $D=10^2$ (circles). The horizontal black lines indicate the equilibrium value $= (V_-,V_0,V_+)/D$.  }
\end{figure}

First, to get a feeling for the average dynamics, we consider the time evolution of \new{the macrostate distribution $p_x(t) = \lr{\psi(t)|\Pi_x|\psi(t)}$} in Fig.~\ref{fig dynamics} \new{for two different Hilbert space dimensions. In Fig.~\ref{fig dynamics}(a) we start from the nonequilibrium initial condition $[p_-(0),p_0(0),p_+(0)] = (1,0,0)$. For $D=10^4$ (solid lines)} we see an exponential (Markovian) decay of the probabilities to their equilibrium value $(V_-,V_0,V_+)/D = (0.2,0.6,0.2)$ (thin horizontal black lines) as predicted by statistical mechanics. The characteristic evolution time scale is set by $\tau$ and equilibration happens roughly for $t\gtrsim7\tau$. \new{For the smaller Hilbert space dimension of $D=10^2$ (circles) a similar tendency but much larger fluctuations are observed, as expected.}

To continue with the many worlds simulation, we now choose an \emph{equilibrium} initial state with $p_x(0) = V_x/D$, for reasons that will also become clear \blue{later}. Naively, one would expect nothing interesting in such a Universe as indicated in Fig.~\ref{fig dynamics}(b): aside from fluctuations, \new{which are exponentially suppressed as a function of the Hilbert space dimension}, nothing seems to happen. But interestingly, a \emph{completely different picture} is possible: even from an equilibrium state many different interesting nonequilibrium universes \new{can} emerge. However, to warrant such a conclusion, we first of all need to ensure that the histories defined by the present coarse-graining are \new{decoherent}, i.e., they obey the DHC condition as discussed in the previous section. In the following, the evaluation of the DF is done for constant time intervals $t_{k+1}-t_k$ equal to the nonequilibrium relaxation time scale $\tau$ defined in Eq.~(\ref{eq tau}).

\begin{figure}[t]
 \centering\includegraphics[width=0.49\textwidth,clip=true]{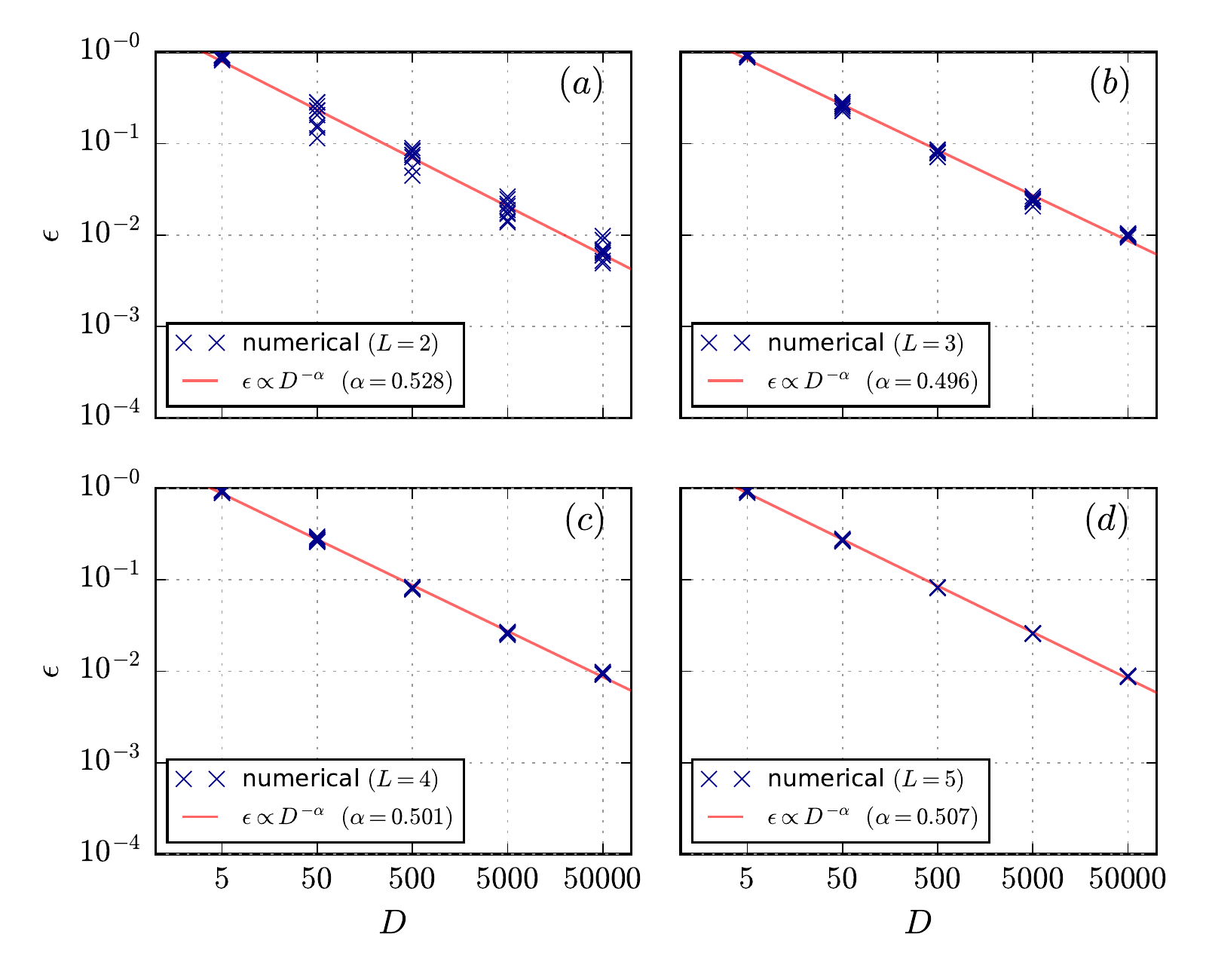}
 \label{fig scale avg eq}\vspace{-0.5cm}
 \caption{Plot of $\epsilon$, the average violation of decoherence defined in Eq.~(\ref{eq epsilon avg}), in the weak coupling regime as a function of the Hilbert space dimension for histories of length $L=2$ (a), $L=3$ (b), $L=4$ (c) and $L=5$ (d). Blue crosses mark results for a single realization of the random matrix interaction and the random initial equilibrium state. Red solid lines fit a scaling law of the form $D^{-\alpha}$ to their averages and the value of $\alpha$ is indicated in the legend inset. Note the double logarithmic scale. }
\end{figure}

\begin{figure}[t]
 \centering\includegraphics[width=0.49\textwidth,clip=true]{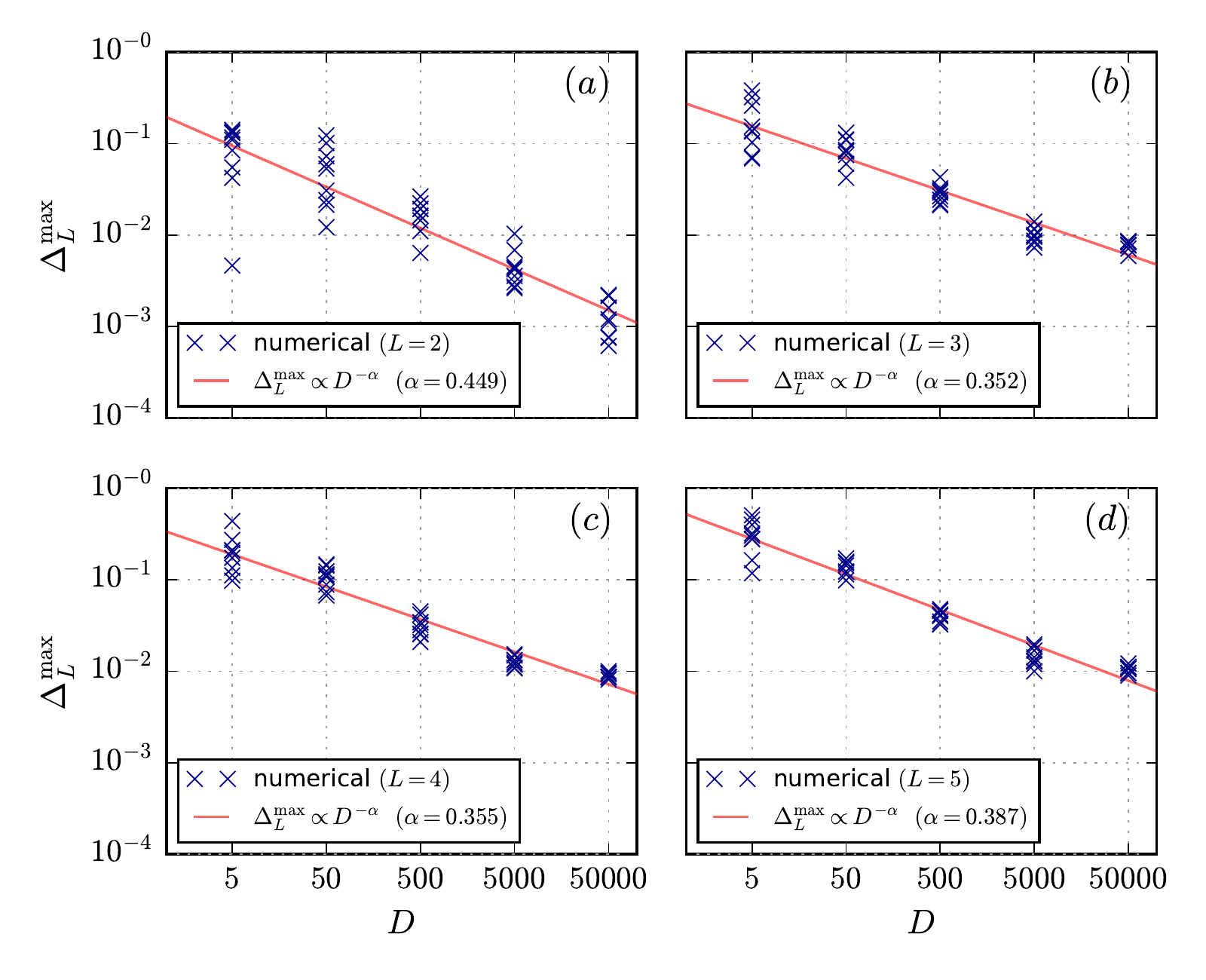}
 \label{fig scale max eq}\vspace{-0.5cm}
 \caption{Plot of $\Delta_L^\text{max}$, the worst case violation of decoherence defined in Eq.~(\ref{eq TD max}), with otherwise the same characteristics as in Fig.~\ref{fig scale avg eq}.}
\end{figure}

The emergence of \new{decoherence} is shown in Figs.~\ref{fig scale avg eq} and~\ref{fig scale max eq}. First,
Fig.~\ref{fig scale avg eq} plots the average off-diagonal elements of the normalized DF as defined in
Eq.~(\ref{eq epsilon avg}) for histories of lengths $L\in\{2,3,4,5\}$ as a function of the Hilbert space dimension $D$.
Each marker in Fig.~\ref{fig scale avg eq} corresponds to one particular random matrix Hamiltonian with one particular
Haar random initial equilibrium state defined in Eq.~(\ref{eq initial state}). In total, we consider three different
realizations of the random matrix Hamiltonian and, for each, three different realization of the initial state, amounting
to $3\times3=9$ different realizations. To extract the overall trend, we average the markers and fit a scaling law of the
form $D^{-\alpha}$ (solid red line).

Two important observations are contained in Fig.~\ref{fig scale avg eq}. First, the different datapoints corresponding to different realizations lie very close to each other, i.e., each realization gives rise to a similar value of $\epsilon$. This indicates \emph{typicality}: most random matrix interactions and most initial states give rise to the same behaviour. Therefore, it makes sense to fit a scaling law $D^{-\alpha}$ to the averaged datapoints. Interestingly, this scaling law, which we extract by varying $D$ over four orders of magnitude from $D=5$ to $D=50000$, is very close to $1/\sqrt{D}$. \new{This scaling suggests that the conditional states $|\psi(\bb x)\rangle$, after normalization, can be expressed as $\sum_i c_i(\bb x)|i\rangle/\sqrt{D}$ for some fixed basis $|i\rangle$. Here, the $c_i(\bb x)$ are random zero-mean-unit-variance coefficients assumed \emph{independent} of $c_i(\bb y)$ for $\bb x\neq\bb y$ because then a random walk argument suggests
\begin{equation}
 \lr{\psi(\bb y)|\psi(\bb x)} \sim \frac{1}{D} \sum_i c_i^*(\bb y) c_i(\bb x) \sim \frac{1}{\sqrt{D}}.
\end{equation}
While it seems plausible, it remains an open question} whether this inverse square root dependence holds for all weakly perturbed random matrix theory models and Haar random initial equilibrium states. In any case, the scaling law suggests that quantum effects are exponentially suppressed as a function of the particle number $N$. Moreover, these conclusions are robust under changing the length $L\in\{2,3,4,5\}$ of the histories. \new{Partial results for very long histories have been also recently obtained by two of the authors~\cite{StrasbergSchindlerArXiv2023}.}

The same conclusions are also reached in Fig.~\ref{fig scale max eq}, where we plot the maximum value of the trace
distance defined in Eq.~(\ref{eq TD max}). Also for this more rigorous quantifier of \new{decoherence} we observe
typicality and an exponential suppression of quantum effects as a function of the particle number $N$. However,
since $\Delta_L^\text{max}$ is determined by the worst case scenario, stronger fluctuations are visible compared to
$\epsilon$, where the average tames fluctuations. Moreover, the exponent $\alpha\approx0.4$ is slightly smaller than
the exponent extracted in Fig.~\ref{fig scale avg eq}. Nevertheless, Fig.~\ref{fig scale max eq} shows
that also statistical outliers, which could be potentially detected by a clever and patient experimentalist, do not
corrupt our conclusions. We conclude that the emergence of \new{decoherence} seems to be a stable and robust phenomenon.

\begin{figure}[t]
 \centering\includegraphics[width=0.40\textwidth,clip=true]{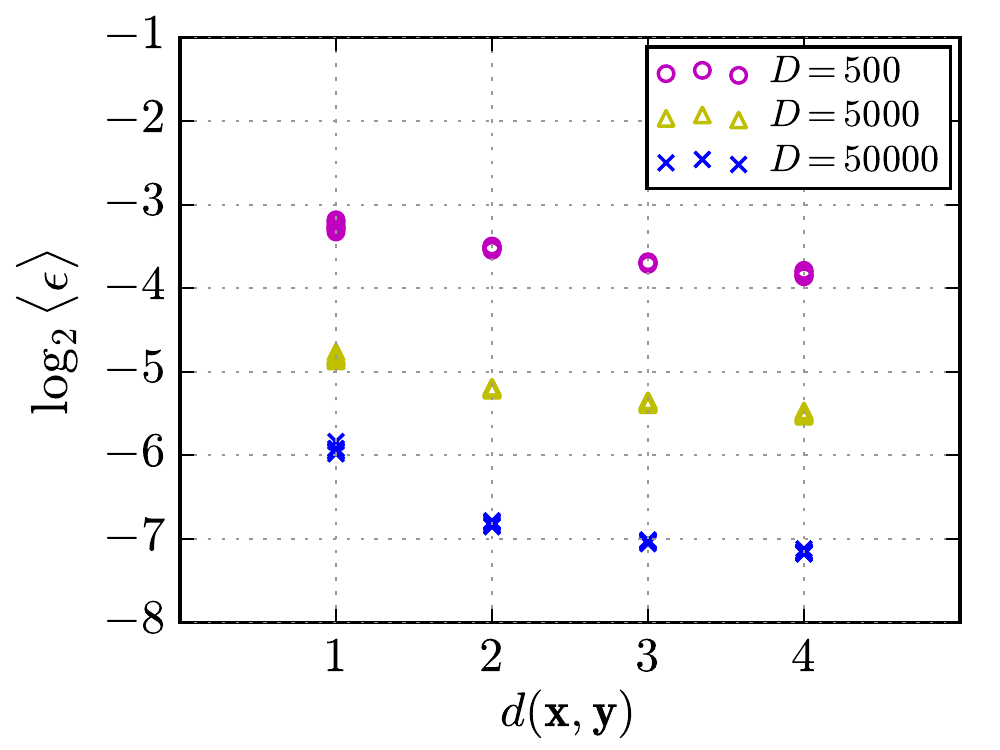}
 \label{fig distances}
 \caption{\blue{Average decoherence $\lr{\epsilon}$ as a function of the history distance $d(\bb x,\bb y)$. Here, $\lr{\epsilon} = \sum_{d'=d(\bb x,\bb y)}\epsilon(\bb x;\bb y)/\#(d')$ sums the normalized DF in Eq.~(\ref{eq approximate decoherence}) over all pairs of histories with a fixed distance $d'=d(\bb x,\bb y)$, divided by the number $\#(d')$ of such pairs. We display results for $D=500$ (magenta circles), $D=5000$ (yellow triangles) and $D=50000$ (blue crosses) for $L=5$ and for the same nine realizations as in Figs.~\ref{fig scale avg eq} and~\ref{fig scale max eq}.} }
\end{figure}

\blue{It is intriguing to ask whether histories or branches that look more distinct from a macroscopic point of view are characterized by stronger decoherence. Specifically, let us define the Hamming distance $d(\bb x,\bb y)$ between two histories $\bb x$ and $\bb y$ as the number of labels where they differ. For instance, for $\bb x = (0,+,0,-,0)$ and $\bb y = (0,-,+,0,0)$ we have $d(\bb x,\bb y) = 3$. Even though, to the best of our knowledge, this question has never been asked before, it seems intuitive to expect that more distinct histories show stronger decoherence. This intuition is indeed confirmed in Fig.~\ref{fig distances}, even though the dependence on the distance $d(\bb x,\bb y)$ is rather mild: an approximate twofold increase in decoherence when going from $d(\bb x,\bb y) = 1$ to $d(\bb x,\bb y) = 4$. However, in more realistic scenarios one would need to take into account much longer histories. Moreover, it seems that the differences are more pronounced for larger Hilbert space dimensions $D$. Unfortunately, current numerical limitations do not allow us to draw strong conclusions from our data.}

\begin{figure}[t]
 \centering\includegraphics[width=0.49\textwidth,clip=true]{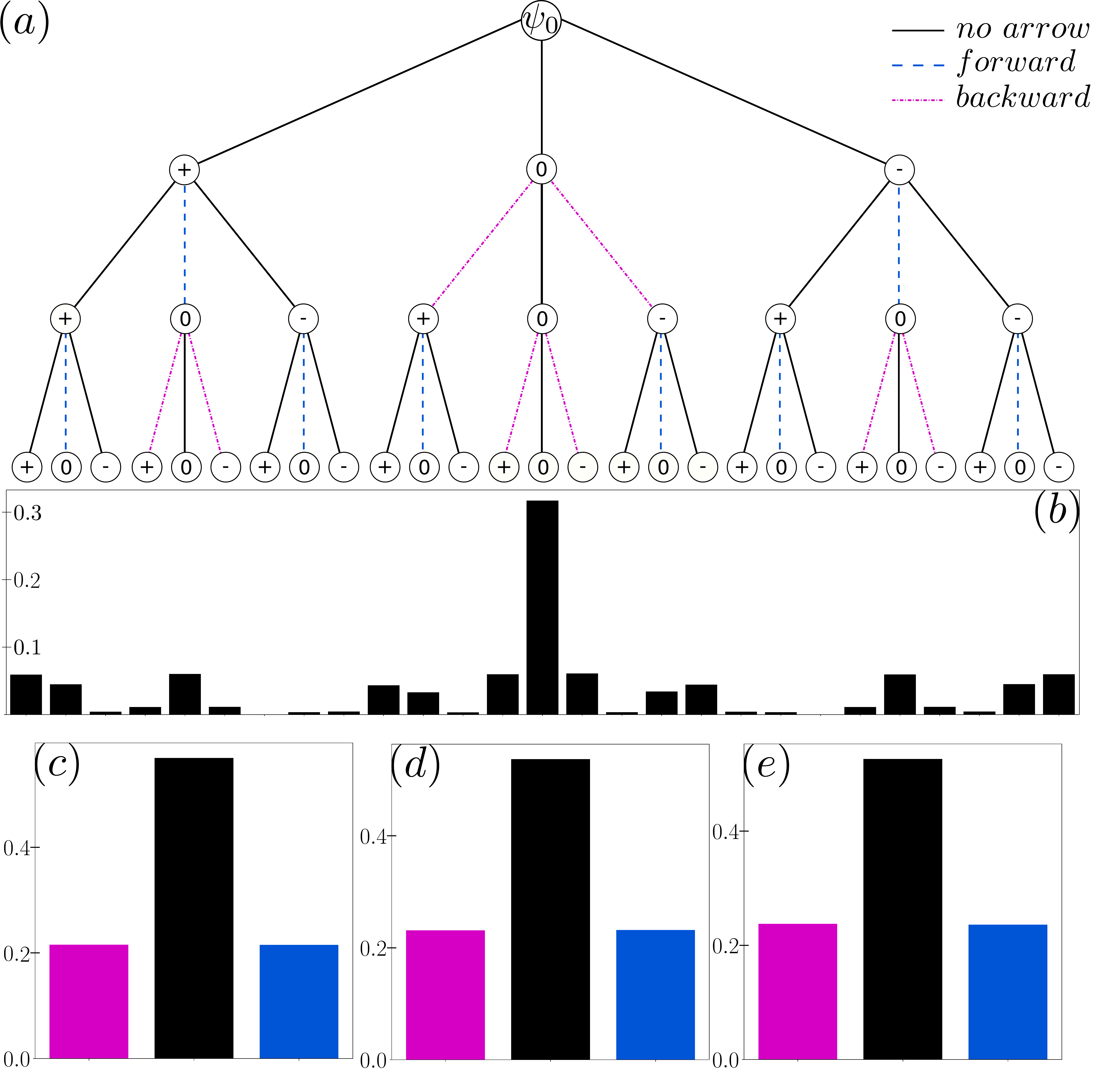}
 \label{fig tree}
 \caption{(a) Depiction of the branching structure of the Multiverse with respect to the initial time $t_0$ and with respect to our chosen coarse-graining. (b) Histogram for the probabilities of different histories. For $L=3$ (c), $L=4$ (d) and $L=5$ (e), respectively, we plot the probability of universes with different (net) arrows of time as explained in the main text. All probabilities were obtained for a single realization of the interaction Hamiltonian (at weak coupling and for $D=10000$) and initial equilibrium state. }
\end{figure}

The overall picture that emerges from our model is depicted in Fig.~\ref{fig tree}. Figure~\ref{fig tree} (a) shows a branching tree structure with respect to histories defined by energetic macrostates labeled by $x\in\{-,0,+\}$ for $L=3$ time steps. Importantly, we can now reason about these histories using a \emph{classical} ontological model for sufficiently large $D$. To facilitate talking about it, we use Boltzmann's entropy concept $S_B(x) = k_B\ln V_x$. We label transitions of the system from a lower to a higher Boltzmann entropy state (i.e., from $+$ or $-$ to $0$) as ``forward'' transitions (blue dashed lines) to indicate that they comply with the conventional \blue{second law-like} arrow of time when $t_0$ denotes the initial time. However, ``backward'' transitions (pink dash-dotted lines) from a higher to a lower entropy state (i.e., from $0$ to $+$ or $-$) are also possible. Finally, transitions involving no change in Boltzmann entropy are labeled by ``no arrow'' (black lines). The Multiverse thus consists of different histories or branches that do not interfere and that describe universes with different \blue{entropic} arrows of time (including the possibility of universes with no arrow of time and both arrows of time) determined by the question whether heat flows from hot to cold or vice versa. \blue{Note that these entropic arrows of time should be distinguished from the ``branching arrow of time'' (i.e., the direction in which the number of branches increases). The branching arrow is pure convention and unrelated to the perceived entropic arrow of time within the universes.}

For one realization of the Hamiltonian and one realization of the initial state we depict the probability for each 
history, which can be now interpreted classically, by the histogram below the tree-like branching structure in
Fig.~\ref{fig tree} (b). As a consistency check, we immediately see that the history $\bb x=(0,0,0)$, in which the
universe resides in the dominant equilibrium macrostate $x=0$, is the most probable one. Note that the probability
for this history is different from the probability $(V_0/D)^3 = 0.6^3= 0.216$ that one would expect by sampling the
histories at time intervals comparable to the equilibration time. The time steps $\tau$ we use are clearly in the nonequilibrium regime.

An important consistency check is to ask about the probability of forward or backward arrows of time. Since the Schr\"odinger equation obeys time reversal symmetry and since the initial equilibrium state that we chose also does not introduce any time asymmetry, we should expect to get a time symmetric answer in this case. The answer is shown in Fig.~\ref{fig tree} (c), (d) and (e) where we plot the probability for a forward arrow of time (blue), no arrow of time (black) and a backward arrow of time (pink) for histories of length $L=3$, $L=4$ and $L=5$, respectively. Here, histories with multiple arrows of time contribute according to their ``net'' arrow of time, i.e., the history $\bb x=(0,+,0)$ has no overall arrow of time. Owing to the fact that we have only three macrostates, there can be no histories with two or more net forward or backward arrows of time. \blue{Thus, our model only exhibits ``mini'' entropic arrows of time, unable to account for our arrow of time in reality. Clearly, by including more than three energy windows in the coarse graining, we could observe longer arrows of time in this model, but their probability (when starting from an equilibrium state) would be very small albeit non-zero. This is the (in)famous Boltzmann brain paradox~\cite{AlbrechtSorboPRD2004, CarrollInBook2020, MuellerQuantum2020}. In any case,} we see that the arrows are indeed symmetrically distributed in our model as it should be. Some non-visible small random fluctuations (in the third significant digit) are due to the fact that a random initial state can still have a small preference to evolve into a higher or lower entropy region, i.e., it is not perfectly symmetric under time reversal. The histograms in Fig.~\ref{fig tree} are based on a single realization of the Hamiltonian and initial state, but we again found this behaviour to be typical.

\begin{figure}[t]
 \centering\includegraphics[width=0.49\textwidth,clip=true]{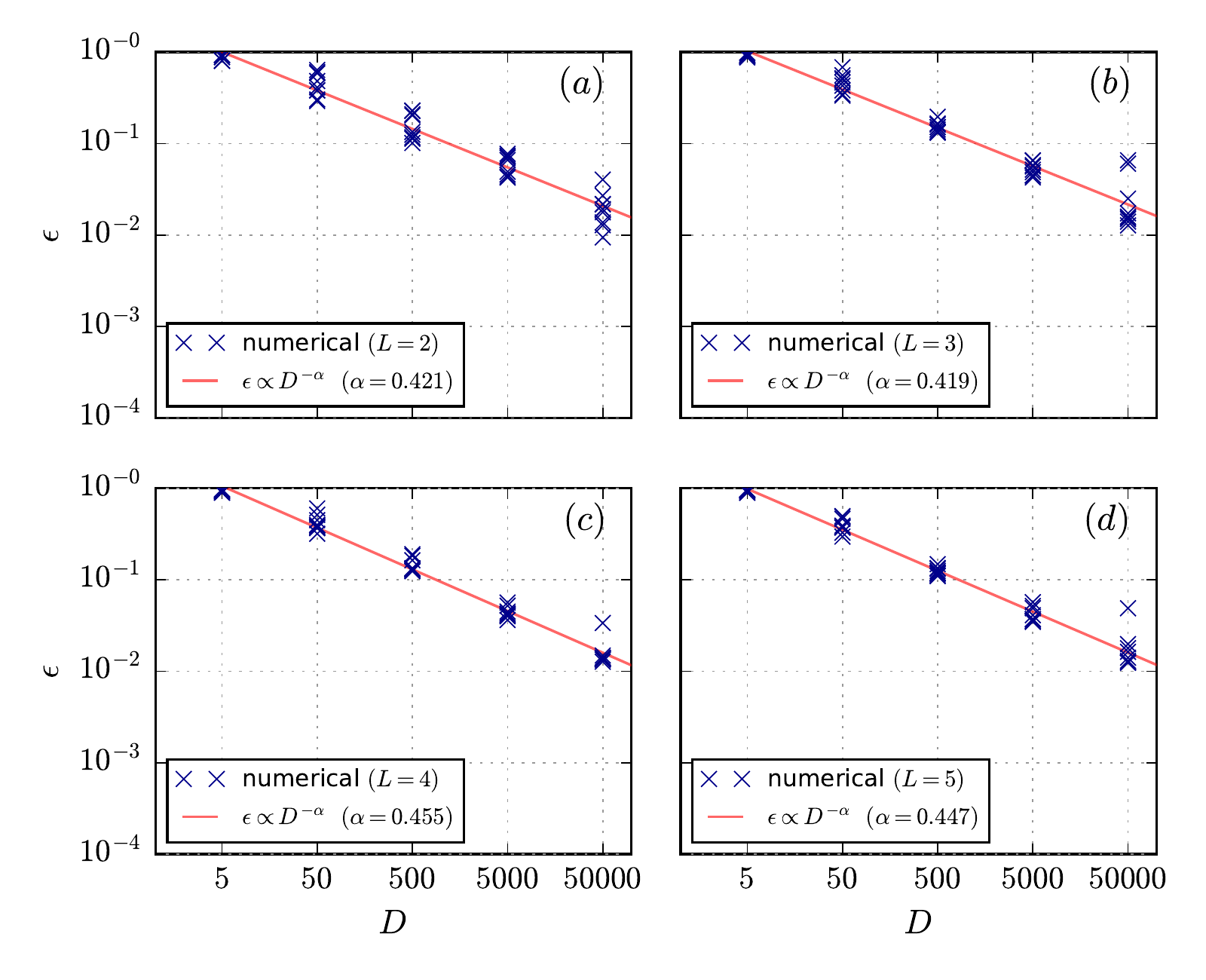}
 \label{fig scale avg eig}\vspace{-0.5cm}
 \caption{Plot of $\epsilon$ for a random initial energy eigenstate. Blue crosses refer to three different randomly chosen eigenstates repeated for three different realizations of the random matrix coupling, i.e., nine realizations in total. Other details are as in Fig.~\ref{fig scale avg eq}. }
\end{figure}

We continue by challenging our model. This is first done by considering an atypical initial state in form of a
randomly selected energy eigenstate $|k\rangle$ of the full Hamiltonian. From the eigenstate thermalization
hypothesis~\cite{DeutschPRA1991, SrednickiPRE1994, SrednickiJPA1999, DAlessioEtAlAP2016, DeutschRPP2018,
ReimannDabelowPRE2021} it is known that these states give rise to probabilities $p_x = \lr{k|\Pi_x|k}$ very close to
the thermal (microcanonical) prediction $V_x/D$, i.e., a picture similar to Fig.~\ref{fig dynamics} (b). However,
it is currently not known whether multi-time correlation functions such as the DF display the same behaviour for 
energy eigenstates and Haar randomly sampled equilibrium states.

\begin{figure}[t]
 \centering\includegraphics[width=0.49\textwidth,clip=true]{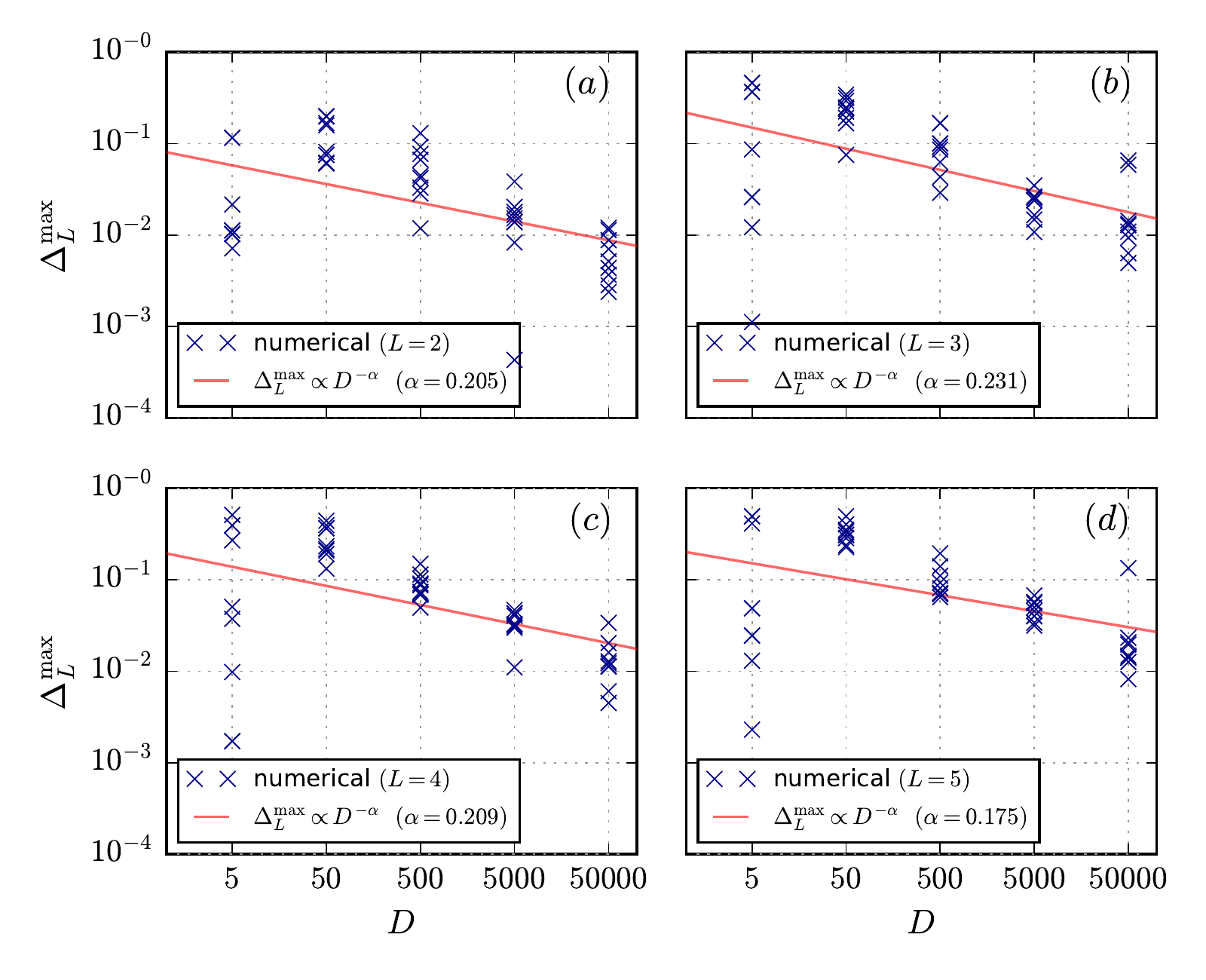}
 \label{fig scale max eig}\vspace{-0.5cm}
 \caption{Plot of $\Delta_L^\text{max}$ with otherwise the same characteristics as in Fig.~\ref{fig scale avg eig}. }
\end{figure}

The results for the emergence of \new{decoherence} with initial energy eigenstates are shown in Figs.~\ref{fig scale avg eig} and~\ref{fig scale max eig}. We again observe a rather typical behaviour \blue{in Fig.~\ref{fig scale avg eig}} and a scaling law suggesting the exponential suppression of quantum effects as a function of the particle number. However, compared with Figs.~\ref{fig scale avg eq} and~\ref{fig scale max eq} the variance in the datapoints is appreciably larger, i.e., there is less typicality. Moreover, the exponents are different. For $\epsilon$ we find $\alpha \approx 0.44$, which is close to the previous value of $\alpha\approx0.5$. Instead, for $\Delta_L^\text{max}$ we find $\alpha\approx0.2$, which is half of the value we found for Haar random initial equilibrium states. \blue{Moreover, $\Delta_L^\text{max}$ fluctuates strongly for different realizations such that the fitted scaling law (red solid line) provides only a rough orientation. Thus,} the preliminary conclusions that we can draw from these data is that \new{correlation functions of} energy eigenstates behave differently from Haar random equilibrium states (which will typically overlap with many energy eigenstates)\new{, which is an interesting result in its own right. In particular, Albrecht \emph{et al.} recently introduced an ``eigenstate einselection hypothesis'' (Appendix B in Ref.~\cite{AlbrechtBaunachArrasmithPRD2022}), where they claim that the ``general features of the consistent histories quantities are unchanged'' for energy eigenstates. On a quantitative level we see here clear evidence for the opposite, but} from a qualitative point of view we observe the same: the difference between $D^{-0.4}$ and $D^{-0.2}$ might not matter in practice, say, for $D\ge 10^{100}$. Thus, the emergence of \new{decoherence} as studied here seems to be a rather robust phenomenon if we restrict the attention to slow and coarse observables of a non-integrable quantum many-body system.

\new{We remark that we have also studied the case of perturbed projectors of the form $V_\delta \Pi_x V_\delta^\dagger$, where the unitary $V_\delta = e^{iA\delta}$ rotates the projectors using a Hermitian matrix $A = A^\dagger$. However, for various $A$ we found quantitatively similar results (apart from fluctuations).\footnote{\new{We checked three cases: $A_{ij} = \delta_{i\pm1,j}$ and $A_{ij} = \delta_{i+j,D+1}$ (with respect to the eigenbasis of the unperturbed Hamiltonian) and $A$ randomly drawn according to the same distribution as $H_I$.}} Therefore, and also to keep the manuscript concise, we do not display these results here, but we believe a systematic study is an interesting task for the future.}

\new{Instead, we find it important to illustrate that} the emergence of \new{decoherence} is \emph{not} a universal phenomenon valid for all observables. To demonstrate this, we consider the case of two strongly coupled systems exchanging energy. To this end, we increase $\lambda$ to $10\lambda$ such that the right hand side of the first equation in Eq.~(\ref{eq conditions epsilon}) equals one. Due to the strong interaction, local energy exchanges will now happen quickly. In some sense, it is no longer meaningful to talk about the local energies of $A$ and $B$ since the local energy levels of systems $A$ and $B$ will strongly hybridize to form new levels.

\begin{figure}[t]
 \centering\includegraphics[width=0.49\textwidth,clip=true]{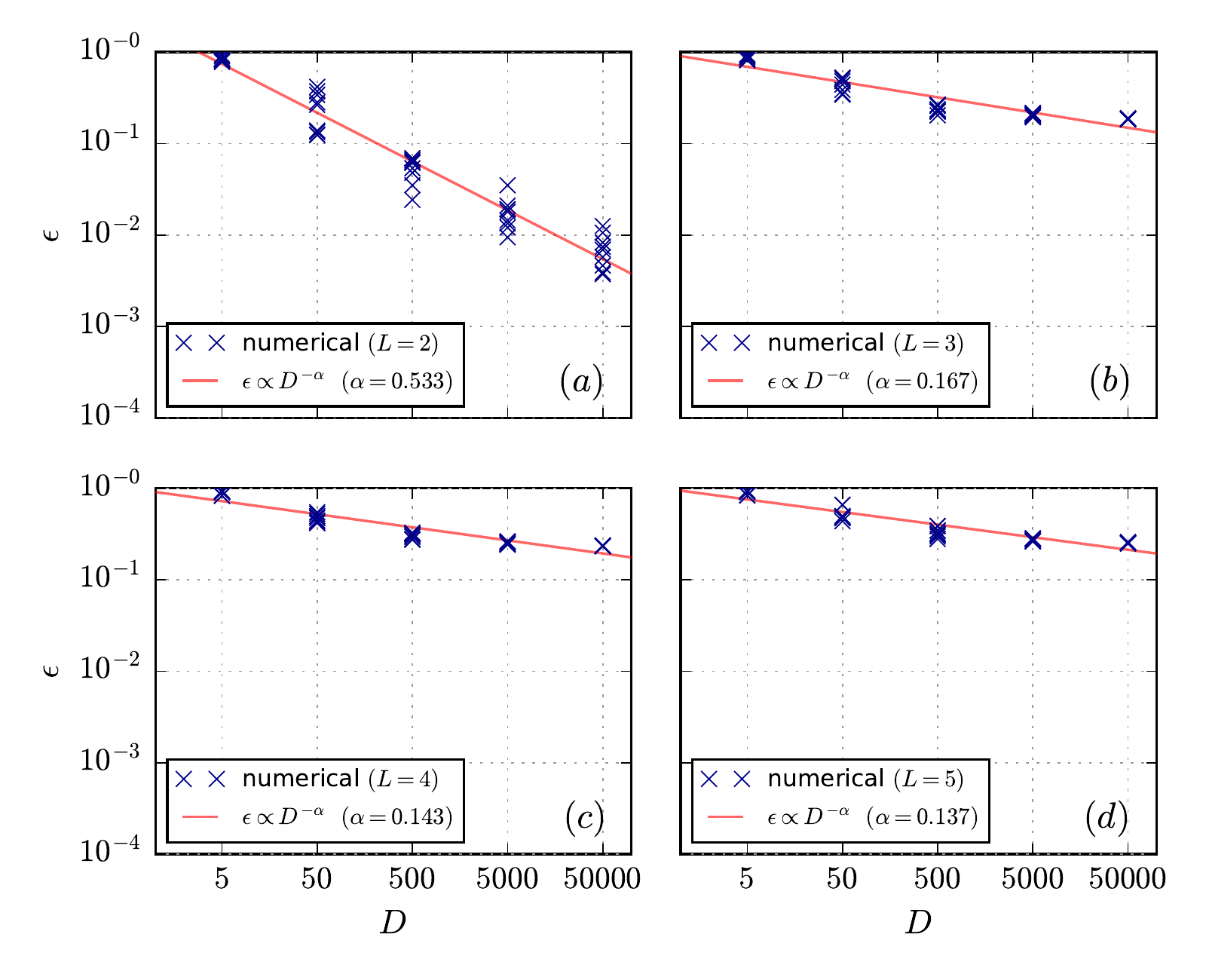}
 \label{fig scale avg eq fast}\vspace{-0.5cm}
 \caption{Plot of $\epsilon$ at strong coupling with otherwise the same characteristics as in Fig.~\ref{fig scale avg eq}. }
\end{figure}

\begin{figure}[t]
 \centering\includegraphics[width=0.49\textwidth,clip=true]{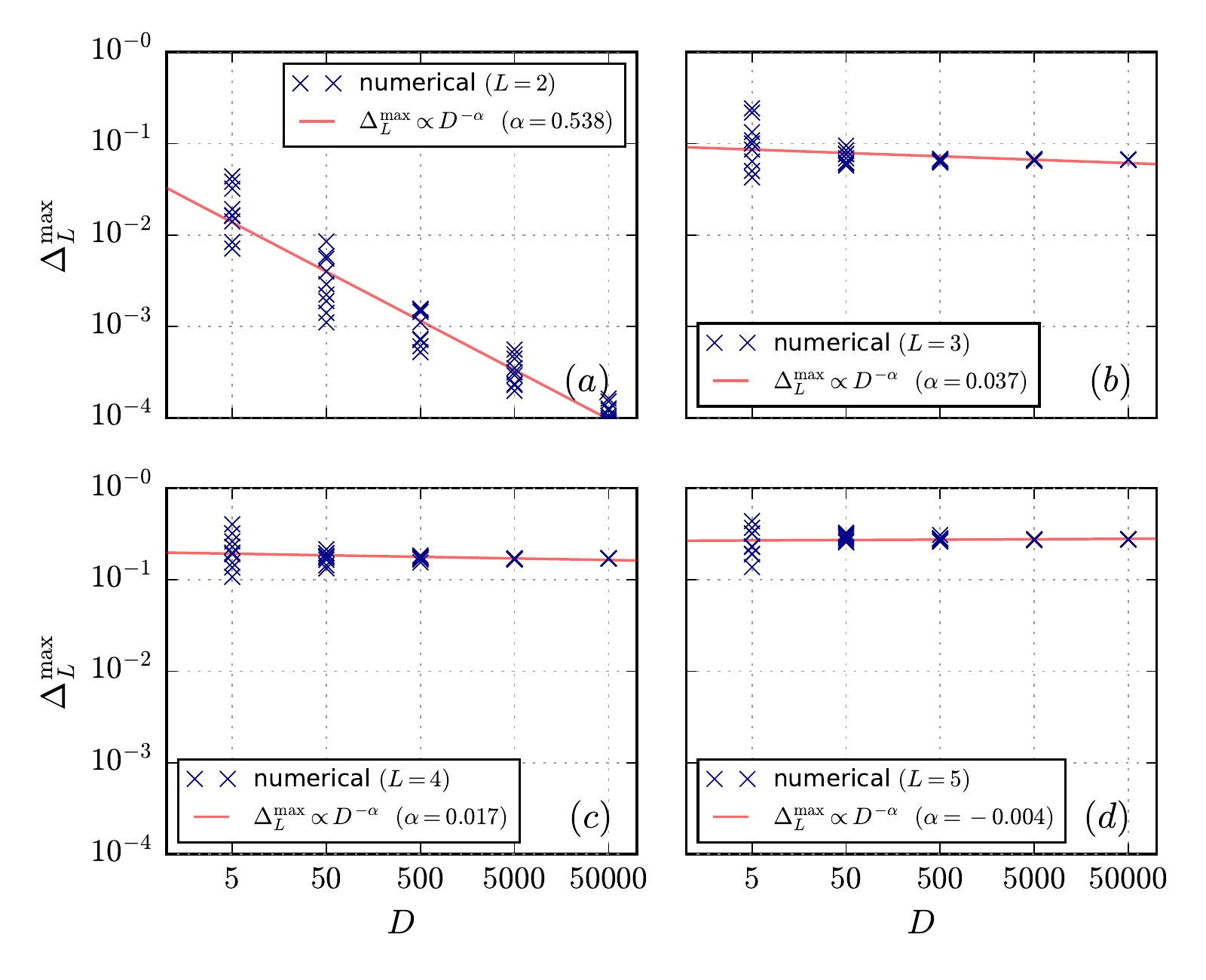}
 \label{fig scale max eq fast}\vspace{-0.5cm}
 \caption{Plot of $\Delta_L^\text{max}$ at strong coupling with otherwise the same characteristics as in Fig.~\ref{fig scale max eq}. }
\end{figure}

\blue{While the considered observable has the same coarseness, we now observe a quite} different behaviour of \new{decoherence} in Figs.~\ref{fig scale avg eq fast} and~\ref{fig scale max eq fast}. While typicality still holds well, \blue{a much milder exponential suppression of quantum effects is observed for $\epsilon$ and} histories of length $L\ge3$, \blue{with a} scaling exponent around $\alpha = 0.15$ (Fig.~\ref{fig scale avg eq fast}). \blue{Even more drastic changes appear for $\Delta_L^\text{max}$ (Fig.~\ref{fig scale max eq fast}) with an exponent $\alpha\approx 0$ for $L=5$ (indicating perhaps a power-law suppression or no suppression at all)}. Recalling that we have to base our conclusions on a finite (and rather small) amount of samples, a clear-cut conclusion seems not possible: exponential suppression of quantum effects, which requires $\alpha>0$, seems possible but is not warranted, and it is certainly much weaker than in the weak coupling (slow observable) regime.

Notably, exponential suppression of quantum effects (with a large exponent $\alpha\approx0.5$) still
holds for the shortest length of histories with $L=2$. This seems counterintuitive, but it can be explained by
recalling that we have chosen a Haar random initial equilibrium state of the form~(\ref{eq initial state}) with
$p_x(0) = V_x/D$ in Figs.~\ref{fig scale avg eq fast} and~\ref{fig scale max eq fast}.
The two-time DF then only probes quantum features of a two-time equilibrium correlation function. They are \new{decoherent} because, first, $\mathbb{E}[\Pi_{x_0}|\psi(t_0)\rl\psi(t_0)|\Pi_{y_0}] \sim \delta_{x_0,y_0}$ where
$\mathbb{E}[\dots]$ denotes a Haar random average, and second, typicality implies that most pure states
$|\psi(t_0)\rangle$ behave similarly. Indeed, for this situation an analytical proof of \new{decoherence} has been given
in Ref.~\cite{StrasbergEtAlPRA2023}.

We end this central part of the manuscript by pointing out that we have tested the robustness of our conclusions
in various further situations for an initial Haar random equilibrium state. For instance, we considered random time 
spacings $t_{k+1}-t_{k}$ chosen uniformly either from $[0,\tau]$ or $[\tau,2\tau]$ (instead of the here considered 
constant time spacing of $t_{k+1}-t_{k} = \tau$), we considered random instead of equally spaced energies in the 
diagonal Hamiltonian blocks of Eq.~(\ref{eq block H}), we used the Gaussian unitary ensemble instead of the Gaussian 
orthogonal ensemble for the off-diagonal blocks of Eq.~(\ref{eq block H}), and we also plotted other quantifiers that 
one could derive from the DHC~(\ref{eq decoherence condition}). In \emph{all} these cases, we have observed the 
\emph{same} qualitative behaviour, i.e., the typicality of a decay law of the form $D^{-\alpha}$. We do not show these 
results here to keep the manuscript at a reasonable length.

\section{\blue{Set Selection Problem}}
\label{sec consequences}

The set selection problem has been used to criticize the histories formalism by pointing out that the DHC is incomplete~\cite{PazZurekPRD1993, DowkerKentPRL1995, DowkerKentJSP1996, KentPS1998, ZurekRMP2003, RiedelZurekZwolakPRA2016, SchlosshauerPR2019, ZurekEnt2022}. The basic reasoning behind the criticism is the following. Consider for fixed $L$ and $M$ the manifold $\texttt{H}$ of all possible projectors that can be used to define histories $\bb x$ (of very different physical meaning). The dimension of this manifold scales like $D^{2L}$, whereas the number of constraints imposed by the DHC in Eq.~(\ref{eq decoherence condition}) scales like $M^{2L}$. Thus, the submanifold $\texttt{DH}$ of \emph{decoherent} histories has a dimension that scales like $D^{2L} - M^{2L}$, which (even though of measure zero with respect to $\texttt{H}$) is enormous for $D\gg M$. Hence, which of the many physically very distinct histories should one use? Note the close similarity to the preferred basis problem.

\blue{We here discuss the set selection problem based on the robust numerical findings from above (for further discussion see also Ref.~\cite{Griffiths2019}). In particular, and contrary to previous claims, we suggest that there simply is} \emph{no} set selection problem \emph{if} one focuses on slow and coarse observables of many-body systems---that is: situations relevant to human perception---and the remaining \blue{freedom in the choice of observable} is actually important \blue{for the ability of different observers to agree}.

\begin{figure}[t]
 \centering\includegraphics[width=0.49\textwidth,clip=true]{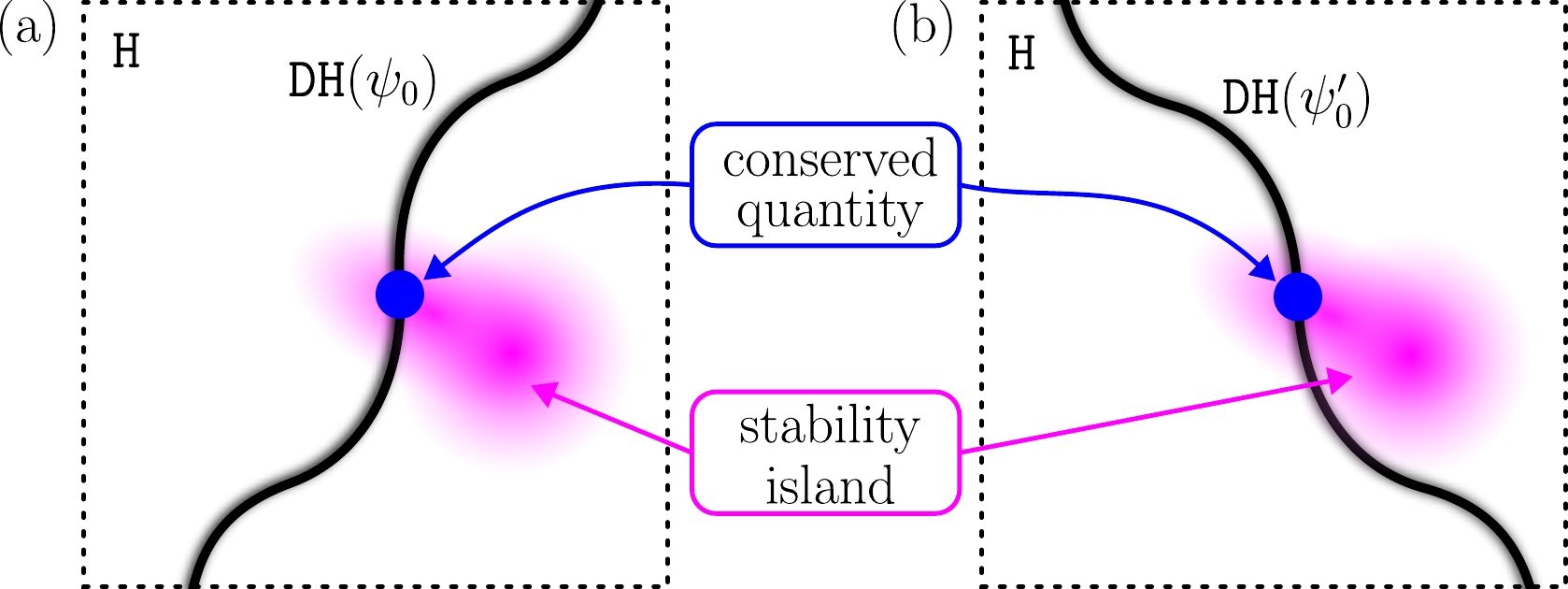}
 \label{fig set selection problem}\vspace{-0.5cm}
 \caption{The square depicts the manifold $\texttt{H}$ of all histories and the shade indicates the amount of decoherence (white regions correspond to no decoherence). The dark black line depicts the submanifold $\texttt{DH}$ of exactly decoherent histores, which is very sensitive to different initial states $\psi(t_0)$ (a) and $\psi'(t_0)$ (b). All $\texttt{DH}(\psi_0)$ intersect for histories of conserved quantities (blue dot, exaggerated in size). Most important to our argument is the stability island (pink shaded region) of approximately decoherent histories, which remains (almost) unchanged for (a) and (b). }
\end{figure}

The \blue{key to our argument} is to demand a certain stability or robustness of decoherence with respect to the initial state $|\psi(t_0)\rangle$ and times $t_k$ \blue{and to focus on approximate instead of exact decoherence, which is sufficient for all practical purposes}. To \blue{recall our results above}: we found approximate decoherence without exception for slow and coarse observables for a large set of initial states, times and Hamiltonians. In contrast, it seems likely that the vast majority of decoherent histories in $\texttt{DH}$ is \emph{not} robust in this sense. Instead, it likely depends sensitively on, e.g., $|\psi(t_0)\rangle$, as sketched in Fig.~\ref{fig set selection problem}.

For instance, one example is a coarse-graining $\{\Pi_x(t_k)\}$ at time $t_k$ with one projector, say the first for $x=1$, equal to $\Pi_{1}(t_k) = |\psi(t_k)\rl\psi(t_k)|$. Here, $|\psi(t_k)\rangle = U_{k,0}|\psi(t_0)\rangle$ is the unitarily evolved initial state. This coarse-graining satisfies by construction the DHC, but it is very sensitive to $\psi(t_0)$ and $t_k$: changing them while leaving the projectors fixed will quickly destroy decoherence.

One might object that those examples are nevertheless legitimate, but there are fundamental practical obstacles. First, it is unclear how humans should get access to a projector of the form $|\psi(t_k)\rl\psi(t_k)|$ using earthly instruments. Second, \blue{if the MWI is correct, then knowledge of the global $|\psi(t_k)\rangle$ is impossible since observers are limited to their own branch $|\psi(\bb x)\rangle$. Finally,} proper relativistic considerations teach us that at any given time $t$ we can only access part of the universal wave function $\rho_\text{accessible}(t) = \mbox{tr}_\text{inaccesible}\{|\psi(t)\rl\psi(t)|\}$. Thus, even if we assume we find (by some magic instrument) a strange decoherent history depending on $\rho_\text{accessible}(t)$, decoherence would likely be destroyed in the next second once a new photon from an hitherto unobserved star reaches us.

Admittedly, this is not a rigorous argument showing that most of the histories in \texttt{DH} are of this fragile nature, but one should also admit in defense of this argument that an interesting example of stable and robust decoherent histories (\emph{beyond} what we discussed here) has never been presented. Thus, we conjecture that the set selection problem is solvable by focusing on locally and practically accessible, approximately decoherent histories that are stable and robust.

We further strengthen the argument by pointing out the similarity with debates dating back to Boltzmann, Loschmidt and Zermelo. For a given initial state it is possible to find a continuum of observables that does \emph{not} thermalize or entropies that \emph{violate} the second law, a fact well known in pure state statistical mechanics. But all these artificially constructed counterexamples are of little relevance because the smallest perturbation of the initial state or Hamiltonian causes them to behave in agreement with thermodynamics again.

In addition, we believe that \emph{approximate} (instead of exact) decoherence is key to resolve this problem because approximate decoherence can be very robust, see the ``stability island'' in Fig.~\ref{fig set selection problem}.\footnote{In addition, our scaling law indicates for realistic many-body systems that approximate decoherence becomes practically indistinguishable from exact decoherence.} Indeed, two different observers never perceive exactly the same observable (for instance, by reading of a display at slightly different angles), but both should still have compatible perceptions (this point is also central to quantum Darwinism~\cite{ZurekNP2009, KorbiczQuantum2021, ZurekEnt2022}). Thus, instead of trying to find a single spot in the history space $\texttt{H}$ that is exactly decoherent, it is more important to find the extended region that guarantees robust approximate decoherence. In this respect, we are sceptical about the idea that approximately decoherent histories can be always slightly distorted to become exactly decoherent~\cite{DowkerKentPRL1995, DowkerKentJSP1996, HalliwellPRD2001, HalliwellPRA2005}: not only is it unclear how to define slight distortion, but it also seems unlikely that this distortion will be robust (independent of $\psi_0$ and $t_k$).

Thus, the picture that emerges is that of a robust stability island that describes accessible and almost exactly decoherent histories for large systems. This stability island is itself a huge region but for good reasons. First, it guarantees that different observers can have similar perceptions of classical reality. Second, the set of slow observables in this stability island will approximately commute for large systems, thus allowing one to define \emph{joint} decoherent histories describing, e.g., the position, momentum and energy of a flying stone.\footnote{However, the technical details of how to precisely handle approximately commuting observables are still subject to research~\cite{HastingsCMP2009, OgataJFA2013, HalpernEtAlNatComm2016}.}

We finish with two remarks. First, there is also an unappreciated set selection problem in EID. Gauge symmetries imply that there is a continuum of physically equivalent system-environment tensor product splittings of the \emph{same} physical system~\cite{StokesNazirNC2019, StokesNazirRMP2022}. Each splitting gives rise to a different system-environment Hamiltonian and, consequently, a different pointer basis. Of course, what matters in practice is that the pointer basis is compatible with the experimental measurement procedure, but this resolution of the ``pointer basis selection problem'' is identical in spirit to the resolution proposed here in the histories context.

Second, we repeat that the proposed solution is restricted to slow and coarse observables of many-body systems that approximately decohere. It does not solve the set selection problem if one demands \emph{exact} decoherence. Also other fundamental problems within the histories formalism such as the one debated in Refs.~\cite{KentPRL1997, GriffithsHartlePRL1998, KentPRL1998} are not directly influenced by our argument.

\section{Concluding Perspectives}
\label{sec perspectives}

\subsection{\blue{Summary}}

We presented a direct evaluation of the DF for multiple (up to five) time steps from first principles for a non-trivial example. We rigorously defined quantum interference effects and extracted a scaling law of the form $D^{-\alpha}$ by varying the Hilbert space dimension $D$ over four orders of magnitude. \blue{This was checked for a wide variety of situations and it} provides a firm starting point to quantitatively discuss the DHC.

The resulting picture is that coarse and slow observables of non-integrable many-body systems give rise to a robust and stable form of approximate decoherence. \blue{While our calculations were restricted to a particular model, the success and versatility of random matrix theory suggests that our results are more widely applicable;} see also Refs.~\cite{VanKampenPhys1954, GemmerSteinigewegPRE2014, SchmidtkeGemmerPRE2016, NationPorrasPRE2020, AlbrechtBaunachArrasmithPRD2022, StrasbergEtAlPRA2023, StrasbergSP2023} for supporting evidence.

\blue{Based on this evidence, we suggested that the preferred basis problem of the MWI or the set selection problem of the histories formalism, respectively, is solvable if one is interested in observables relevant to us humans (which are slow and coarse and give rise to a robust and stable form of decoherence) and if one focuses on approximate instead of exact decoherence. Indeed, our results indicate that for such observables the emergence of decoherence is universal and happens for almost all pure states: no product state assumption, no low entropy initial state and no ensemble average was necessary in our approach. Moreover, it seems reasonable to expect (but difficult to prove) that possible ambiguities in the choice of branches/histories become irrelevant as different slow and coarse obervables approximately commute for large quantum systems.}

\new{Importantly,} this \new{picture} does not contradict previous works, but it complements and extends them by providing new perspectives, tools and insights. For instance, our results do not rely on the widely used concepts of EID and quantum Darwinsim, but they are not in conflict with them. \new{Indeed, a $1/\sqrt{D}$ scaling law has been also extracted for a random matrix model in the context of EID~\cite{HeWangPRE2014}.} Moreover, \new{we believe} non-integrability is a key factor, but it \new{is usually} not regarded as such in other works on the quantum-to-classical transition.

\blue{Finally, we also explicitly saw that an equilibrated Multiverse gives rise to branches with locally well-defined entropic arrows of time while overall the Multiverse remains statistically time-symmetric. While our model could be claimed to be unrealistic because of the Boltzmann brain paradox~\cite{AlbrechtSorboPRD2004, CarrollInBook2020, MuellerQuantum2020}, it nevertheless explicitly illustrates two crucial features. First, the branching structure of the Multiverse is pure convention and does not explain our arrow of time or meaningful thermodynamic entropy, contrary to suggestions made in Refs.~\cite{DeutschInBook2010, Aaronson2013}. Instead, from a fundamental point of view it would be preferable to start from a time-symmetric description of the MWI or histories formalism~\cite{AharonovBergmannLebowitzPR1964, IshamJMP1994, IshamLindenPRA1997, GriffithsBook2002, VaidmanInBook2010}. Second, opposite arrows of time can peacefully coexist within the same quantum Multiverse in contrast to opposite arrows of time in space-like separated regions of a classical Multiverse~\cite{SchulmanPRL1999}, which are unstable with respect to the slightest perturbation~\cite{KupervasserNikolicZlaticFP2012}. For further details about the emergence of classicality in time-symmetric situations see Ref.~\cite{AlbrechtBaunachArrasmithPRD2022}. Moreover, that arrows of time can be an emergent concept in a time reversal symmetric Universe was also noted in different contexts, e.g., in Refs.~\cite{SchulmanPRL1999, CarrollChenArXiv2004, BarbourKoslowskiMercatiPRL2014, DeutschAguirreFP2022}.}

\subsection{\blue{Outlook}}

\blue{Our work could stimulate various research directions as it combines tools and concepts related to the quantum-to-classical transition, quantum cosmology, quantum statistical mechanics and quantum stochastic processes.}

\blue{For instance, while we argued that non-integrability could be a key factor,} chaos has been also realized as detrimental for the emergence of classicality within the Wheeler-DeWitt equation due to the breakdown of the WKB approximaton and EID was invoked to remedy for that~\cite{CalzettaGonzalezPRD1995, CornishShellardPRL1998, CalzettaCQG2012}. Within our non-relativistic quantum mechanical model we can not make any direct contribution to this question, but one view suggested by this work is that emergence of classicality is best viewed as a \emph{synergy} of different mechanisms instead of a single all-ruling idea. Indeed, it has been recently pointed out that the emergence of classicality in the early Universe is still an unsolved puzzle~\cite{BerjonOkonSudarskyPRD2021}, and the present perspective might be able to add a piece to it.

\blue{Moreover,} it would be desirable to get a better analytical understanding of the DHC. The results reported in Refs.~\cite{StrasbergEtAlPRA2023, StrasbergSP2023} were restricted to three-time correlation functions and generalizing them to higher orders likely requires novel techniques (however, some progress for arbitrary $L$-time correlation functions in a different context was reported in Refs.~\cite{FigueroaRomeroModiPollockQuantum2019, FigueroaRomeroPollockModiCP2021, DowlingEtAlQuantum2023, DowlingEtAlSPC2023}). Further insights about the time scales at which \new{decoherence} arises are also desirable, and it is in particular also worth to ask about ``recoherence times'': due to Poincar\'e recurrences the emergence of \new{decoherence} in a finite dimensional quantum system can not be a persistent, eternal phenomenon. In particular, it would be intriguing to find out how the recoherence time scales with the number of time steps $L$ or, more generally, with the \emph{net} information aquired along a history. \blue{In addition, it seems that more research is necessary to understand the general consequences of approximate instead of exact decoherence given that the latter is not the rule.}

We further remark that we have used the conventional framework of non-relativistic quantum mechanics and assumed the validity of the Born rule. A significant fraction of research is devoted to interpreting or understanding the origin of the Born rule within the MWI (see, e.g., Refs.~\cite{VaidmanISPS1998, SaundersEtAlBook2010, AguirreTegmarkPRD2011, WallaceBook2012, Vaidman2020, ZurekEnt2022} \new{and references therein). The present approach could also add insights to this debate when considering very long histories for $L\gg1$, as recently studied by two of the authors~\cite{StrasbergSchindlerArXiv2023}.}

Finally, our model calculations presently suggest that for any coarse and slow observable of a non-integrable system every possible classical history is realized. While this suggests the realization of a wide array of histories, it is \emph{not} necessarily true that ``everything that can happen will happen'' (as sometimes portrayed in both scientific and popular accounts of the MWI). For instance, extremely unlikely (sequences of) outcomes might be restricted to very small subspaces because they are highly atypical, but our results indicate that large subspaces are needed for the emergence of classicality\new{, see also Ref.~\cite{StrasbergSchindlerArXiv2023}}. \blue{We also collected some evidence in Fig.~\ref{fig distances} that there is structure in the Multiverse among different branches, but much more work in that direction needs to be done.} Finally, whether all the perceived randomness in our world stems from quantum effects~\cite{AlbrechtPhillipsPRD2014} \new{or not}~\cite{DelSantoGisinPRA2019} \new{remains a fascinating question too}.

\blue{\emph{Note added.}---While this manuscript was under review, evidence appeared that confirms our results for realistic quantum-chaotic many-body systems and suggests that integrable finite-size systems show a much weaker form of decoherence~\cite{WangStrasbergArXiv2024}.}


\subsection*{Acknowledgements}

We are grateful to Anthony Aguirre, Josh Deutsch, Jonathan Halliwell, and Anastasiia Tiutiakina for discussions.
PS is financially supported by ``la Caixa'' Foundation (ID 100010434, fellowship code LCF/BQ/PR21/11840014)
and acknowledges further support from the European Commission QuantERA grant ExTRaQT (Spanish MICINN project
PCI2022-132965), by the Spanish MINECO (project PID2019-107609GB-I00) with the support of FEDER funds, the Generalitat
de Catalunya (project 2017-SGR-1127) and by the Spanish MCIN with funding from European Union NextGenerationEU
(PRTR-C17.I1). JS acknowledges support by MICIIN with funding from European Union NextGenerationEU (PRTR-C17.I1) and
by the Generalitat de Catalunya.



\bibliography{/home/philipp/Documents/references/books,/home/philipp/Documents/references/open_systems,/home/philipp/Documents/references/thermo,/home/philipp/Documents/references/info_thermo,/home/philipp/Documents/references/general_QM,/home/philipp/Documents/references/math_phys,/home/philipp/Documents/references/equilibration,/home/philipp/Documents/references/time,/home/philipp/Documents/references/cosmology,/home/philipp/Documents/references/general_refs}

\end{document}